\documentclass{aa}

\usepackage{graphicx}
\usepackage{color}
\usepackage{amsbsy}
\usepackage{mathrsfs}
\usepackage{mathpazo,bm}
\usepackage{amsmath}
\usepackage{multirow}
\usepackage{wasysym} 

\newlength{\hfwidth}
\newlength{\hfwidthsingle}
\newcommand{\ptderiv}[1]{\frac{\partial #1}{\partial t}}

\newcommand{\Ma}{\mathrm{Ma}}
\newcommand{\Rey}{\mathrm{Re}}
\newcommand{\Kn}{\mathrm{Kn}}
\newcommand{\ttimes}[1]{10^{#1}}
\newcommand{\xtimes}[2]{#1 \times 10^{#2}}

\newcommand{\vt}[1]{\mathbf{#1}}       
\renewcommand{\v}[1]{{\boldsymbol #1}} 

\newcommand{\del}{\v{\nabla}}
\newcommand{\grad}{\del}
\newcommand{\Div}{\del\cdot}
\newcommand{\curl}{\del\times}
\newcommand{\Laplace}{\nabla^2}

\newcommand{\Sun}{\odot}
\newcommand{\Eps}{{\rm Eps}}
\newcommand{\Stk}{{\rm Stk}}
\newcommand{\dv}{\Delta{\v{v}}}
\newcommand{\mearth}{\,$M_{\oplus}$}
\newcommand{\mearthp}{\,$M_{\oplus}$ }

\newcommand{\mps}{m\,s$^{-1}$}

\newcommand{\mdot}{\dot{m}} 

\newcommand{\Eq}[1]{Eq. (\ref{#1})}

\newcommand{\Eqss}[2]{Eqs. (\ref{#1})--(\ref{#2})}
\newcommand{\eq}[1]{\Eq{#1}}

\newcommand{\App}[1]{Appendix~\ref{#1}}

\newcommand{\Figure}[1]{Figure~\ref{#1}}
\newcommand{\Fig}[1]{Fig.~\ref{#1}}
\newcommand{\fig}[1]{\Fig{#1}}

\definecolor{black}{rgb}{0.0,0.0,0.0}
\def\white#1{\textcolor{white}{#1}}
\def\black#1{\textcolor{black}{#1}}

\begin{document}

\title{Planet formation bursts at the borders of the dead zone \\in 2D numerical simulations of circumstellar disks}

\author{
W. Lyra\inst{1},
A. Johansen\inst{2},
A. Zsom\inst{3},
H. Klahr\inst{3}, \&
N. Piskunov\inst{1}
}

\offprints{wlyra@astro.uu.se}

\institute{Department of Physics and Astronomy, Uppsala Astronomical 
Observatory, Box 515, 751\,20 Uppsala, Sweden
\and Leiden Observatory, Leiden University, PO Box 9513, 2300 RA Leiden, The Netherlands
\and Max-Planck-Institut f\"ur Astronomie, K\"onigstuhl 17, 69117 
Heidelberg, Germany}

\date{Received ; Accepted}

\authorrunning{Lyra et al.}
\titlerunning{Planet formation bursts}

\abstract
{As accretion in protoplanetary disks is enabled 
by turbulent 
viscosity, the border between active and inactive (dead) zones constitutes
a location where there is an abrupt change in the accretion flow. The 
gas accumulation that ensues triggers the Rossby wave instability, 
which in turn saturates into anticyclonic vortices. It has been 
suggested that the 
trapping of solids within them leads to a burst of 
planet formation on very short timescales.} 
{We study in the formation and evolution 
of the vortices in greater detail, focusing on the implications for the dynamics 
of embedded solid particles and planet formation.} 
{We performed two-dimensional global simulations of the dynamics of 
gas and solids in a non-magnetized thin
protoplanetary disk with the Pencil code. We used multiple particle species 
of radius 1, 10, 30, and 100\,cm. We computed the particles' 
gravitational interaction by a particle-mesh method, translating the 
particles' number density into surface density and computing the corresponding 
self-gravitational potential via fast Fourier transforms.
The dead zone is modeled as a region of low viscosity. Adiabatic and locally 
isothermal equations of state are used.}
{The Rossby wave instability is triggered under a variety of 
conditions, thus making vortex formation a robust process. Inside 
the vortices, fast accumulation of solids occurs and the particles 
collapse into objects of planetary mass on timescales as  
short as five orbits. Because the drag force is size-dependent, aerodynamical 
sorting ensues within the vortical motion, and the first bound 
structures formed are composed primarily of similarly-sized particles. 
In addition to erosion due to ram pressure, we identify gas tides 
from the massive vortices as a disrupting agent of formed 
protoplanetary embryos. 
We find evidence that the backreaction of the drag force from the 
particles onto the gas modifies the evolution of the Rossby wave 
instability, with vortices 
being launched only at later times if this term is excluded from the 
momentum equation. Even though the gas is 
not initially gravitationally unstable, the vortices can grow to $Q\approx 1$ 
\black{in locally isothermal runs, which halts the inverse cascade of energy towards 
smaller wavenumbers. As a result, vortices 
in models without self-gravity tend to rapidly merge towards a $m$=2 or $m$=1 
mode, while models with self-gravity retain dominant higher order modes (m=$4$ 
or $m$=3) for longer times. Non-selfgravitating disks thus show 
fewer and stronger vortices. We also estimate the collisional velocity history
of the particles that compose the most massive embryo by the end of the 
simulation, finding that the vast majority of them never experienced 
a collision with another particle at speeds 
faster than 1\mps. This result lends further support to
previous studies showing that vortices provide a favorable
environment for planet formation.}} 
{}

\maketitle
\section{Introduction}
\label{sect:introduction}

The ill fate of the building blocks of planets in gaseous disks around 
young stars stands as one of the major unsolved problems in the theory of 
planet formation. Beginning with micron-sized interstellar dust grains, 
coagulation models predict growth to centimeter 
and meter size (Weidenschilling 1980; Dominik et al. 2007) in the denser 
environments of a circumstellar disk. Such bodies, however, are large 
enough to have already decoupled slightly from the sub-Keplerian gas, 
yet still small enough to be subject to a significant gas drag. The resulting 
headwind drains their angular momentum, leading them into spiral 
trajectories towards the star, on timescales as short as a hundred 
years at 1AU (Weidenschilling 1977a). Another acute problem is that 
such bodies have poor sticking properties and a low threshold velocity 
for fragmentation (Chokshi et al. 1993), such that collisions between 
them usually lead to destruction rather than growth (Benz 2000; 
Sirono 2004; Ormel \& Cuzzi 2007). Such problems severely hinder 
growth to km-size by coagulation (Brauer et al.\ 2008a).

In view of these problems, other routes for breaching the meter size 
barrier have been pursued. A distinct alternative is gravitational 
instability of the layer of solids (Safronov 1969, Lyttleton 1972;
Goldreich \& Ward 1973; 
Youdin \& Shu 2002). When the dust aggregates had grown 
to centimeter and meter size the gas drag is reduced and the solids are 
pushed to the midplane of the disk due to the stellar gravity. Although 
such bodies do not have enough mass to attract each other individually, 
sedimentation increases the solids-to-gas ratio by orders 
of magnitude when compared to the interstellar value of $\ttimes{-2}$.
It was then hypothesized (Safronov 1969) that due to the high densities 
of this midplane layer, the solids could collectively achieve critical 
number density and undergo direct gravitational collapse. Such a scenario has 
the advantage of occurring on very rapid timescales, thus avoiding the 
radial drift barrier.

This picture is nonetheless simplistic, in the view that even low levels 
of turbulence in the disk preclude the midplane layer of solids from achieving 
densities high enough to trigger the gravitational instability 
(Weidenschilling \& Cuzzi 1993). Even in the absence of self-sustained 
turbulence such as the one generated by the magneto-rotational instability 
(MRI; Balbus \& Hawley 1991; Balbus \& Hawley 1998), the 
solids themselves can generate turbulence due to the backreaction of the 
drag force onto the gas. Such turbulence can be 
brought about by Kelvin-Helmholtz instabilities due to the vertical 
shear present in the sedimented layer of solids (Weidenschilling 1980; 
Weidenschilling \& Cuzzi 1993; Sekiya 1998; Johansen et al. 2006), or by 
streaming instabilities induced by the radial migration of 
solids particles (Youdin \& Goodman 2005; Johansen et al.\ 2006; Paardekooper
2006; 
Youdin \& Johansen 2007; Johansen \& Youdin 2007). In the turbulent motion, 
the solids are stirred up by the gas, forming a vertically extended layer 
where the stellar gravity is balanced by turbulent diffusion (Dubrulle et al. 1995; Garaud \& Lin 2004).

But if  turbulence precludes direct gravitational collapse through 
sedimentation, it was also shown that it allows for it in an indirect way. 
As solid particles concentrate in high pressure regions
(Haghighipour \& Boss 2003), the solids-to-gas ratio can be enhanced 
in the transient turbulent gas pressure maxima, potentially reaching 
values high enough to achieve gravitational collapse. Numerical calculations 
by Johansen et al. (2007) show that this is indeed the case, with the 
particles trapped in the pressure maxima generated by the MRI collapsing 
into dwarf planets when the gravitational interaction between particles 
is considered. They also show that the MRI is not necessarily needed, 
since the weak turbulence brought about by the streaming instability itself 
can lead to enough clumping under certain conditions. Another way of 
achieving high enough densities for gravitational collapse of the solid layer 
was shown by Rice et al. (2004) and Rice et al. (2006), where meter-sized 
solids concentrate prodigiously in the spiral arms formed in marginally 
gravitationally unstable circumstellar disks.

Such models, however, ignored the possibility of fragmentation of particles 
upon collisions. As the turbulence enhances the velocity dispersion of 
solids, destructive collisions become more likely. Moreover, upon 
destruction, the smaller fragments are tightly coupled to the gas and 
therefore dragged away from the midplane (Johansen et al. 2008), reducing  
the effective amount of solid material available for collapse. Such problem 
is particularly severe in the high mass disks investigated by 
Rice et al. (2004) and Rice et al. (2006), where the typical speeds of the 
boulders upon encounters are comparable to the sound speed.

The fragmentation problem could be avoided if the accumulation of 
solids happened, for instance, within a protective environment where 
the collisional speeds are brought down to gentler values. Anticyclonic vortices 
(Marcus 1990) have been shown to favor planet formation 
(Barge \& Sommeria 1995; Tanga et al. 1996; Bracco et al. 1999; 
Chavanis 2000; Johansen et al. 2004) since, by rotating clockwise in the 
global counterclockwise Keplerian flow, they enhance the local shear 
and induce a net force on solid particles towards their centers. Klahr \& 
Bodenheimer (2006) further argue that anticyclonic vortices 
would be less turbulent than the 
ambient gas, which in turn would lead to velocity dispersions that are low 
enough to prevent fragmentation. Vortices in disks can be the result 
of the baroclinic instability (Klahr \& Bodenheimer 2003; Klahr 2004; 
Petersen et al. 2007), the Rossby wave instability (Lovelace et 
al. 1999; Li et al. 2000; Li et al. 2001) or, perhaps, the MRI (Fromang 
and Nelson 2005).
 
In this paper, we focus on vortices generated by the Rossby wave instability
(RWI),
which is a global instability where azimuthal modes experience growth 
in the presence of local extrema of a quantity 
interpreted as a combination of entropy and potential vorticity. In the 
linear phase, the instability launches inertial-acoustic
waves. The non-linear saturation is achieved when the Rossby waves break 
and coalesce into anticyclonic vortices. It was shown by 
Varni\`ere \& Tagger (2006) that a favorable 
profile of the entropy-modified vorticity naturally arises if the disk 
has a slow-accretion zone, such as in the layered accretion model 
of Gammie (1996). In this model, ionization is provided by collisions
in the hot inner regions, and by cosmic rays in the outer disk where the 
column densities are low (a standard value for the penetration depth of 
cosmic rays is a gas column density of 100 g\,cm$^{\rm -2}$). Throughout 
most of the midplane, however, 
the temperatures are too cold and the column densities are too thick for 
ionization to occur either way. The result is that, when threaded by a weak 
magnetic field, the disk displays MRI-active regions in the ionized layers, 
and a MRI-dead zone in the neutral parts around the 
midplane (Gammie 1996, 
Miller \& Stone 2000; Oishi et al. 2007). Matter flows towards the star 
due to the high turbulent viscosity of the MRI-active layers, but upon 
hitting the border of the dead zone, it reaches a region of slow accretion 
and the flow stalls. However, as the flow proceeds unabridgedly from the outer 
active regions, a surface density maximum forms, which triggers the growth 
of the RWI.

The implications of this scenario for planet formation were first 
explored by Inaba \& Barge (2006), who use the RWI-unstable dead zone model 
of Varni\`ere \& Tagger (2006) to study the accumulation of solids inside the
Rossby vortices. They confirm that the vortices are efficient particle traps, 
since the solids-to-gas ratio was raised by at least one order of magnitude,
modeling the solid phase of the disk as a fluid. Such 
approximation requires that the size of the solid particles be much smaller 
than the gas mean free path. Since in the Minimum Mass Solar Nebula (MMSN; 
Weidenschilling 1977b) at 5.2AU the particles subject to maximum drift have a size comparable to 
the mean free path, the sizes that a fluid approach can handle correspond to too 
strong friction, thus ultimately underestimating the 
trapping performance of the vortical motion. In Lyra et al. (2008b; hereafter 
LJKP08), we 
took the works of Varni\`ere \& Tagger (2006) and Inaba \& Barge (2006) one 
step further by including gravitationally interacting centimeter and meter 
size solids treated as Lagrangian particles. In that Letter, we showed 
that the solids concentrated in the vortices triggered by the RWI rapidly 
reach critical densities and undergo collapse into rocky planets. The 
resulting burst lead to the formation of 20 rocky protoplanetary embryos 
in the mass range 0.1-0.6 \mearth, along with hundreds of smaller bodies 
following a mass spectrum of power law $-2.3$$\pm$$0.2$.

In this paper we further detail the method used in LJPK08, 
also presenting a number of new results. In the following section we present 
the dynamical equations, followed by an in-depth analysis of the vortices 
in Sect.~\ref{sect:vortices}. In Sect.~\ref{sect:embryos} we analyze 
the formation and evolution of the protoplanetary embryos, focusing 
on stability against erosion (Paraskov et al. 2006; Cuzzi et al. 2008) 
and tides from the gas, which we identify as an important disrupting 
agent. In Sect.~\ref{sect:rwi-response} we investigate the response of 
the RWI to effects not considered in the original analysis of Lovelace 
et al. (1999) and Li et al. (2000), and in Sect.~{\ref{sect:limitations}}
we present a discussion of the limitations of the model. A summary and 
conclusions are presented in Sect.~{\ref{sect:conclusions}}. 

\section{The model}
\label{sect:model}
\subsection{Dynamical Equations}
\label{sect:dynamical-equations}

We work in the thin disk approximation, using the vertically integrated
equations of hydrodynamics

\begin{eqnarray}
\ptderiv{\varSigma_g} &=& -\left(\v{u}\cdot\del\right)\varSigma_g -\varSigma_g{\Div\v{u}} + f_D(\varSigma_g) \label{eq:continuity}\\
\ptderiv{\v{u}} &=& -\left(\v{u}\cdot\del\right)\v{u} -\frac{1}{\varSigma_g}\del{P} - \del\varPhi - \frac{\varSigma_p}{\varSigma_g}\v{f}_d  \\\nonumber
& &+ 2\,\varSigma_g^{-1}\,\Div{\left(\nu\varSigma_g \vt{S}\right)} + f_\nu(\v{u},\varSigma_g) \label{eq:Navier-Stokes}\\
\frac{d{\v{x}}_{p}}{dt} &=& \v{v}_p\label{eq:particle-vel}\\
\frac{d{\v{v}}_{p}}{dt} &=& - \del\varPhi + \v{f}_d \label{eq:particle}\\
\varPhi&=&\varPhi_{\rm sg} -\frac{GM_\Sun}{r}  \label{eq:potential}\\
\Laplace\varPhi_{\rm sg} &=& 4{\pi}G\left(\varSigma_g+\varSigma_p\right)\delta(z) \label{eq:poisson}\\
P&=&\varSigma_g c_s^2    \label{eq:state}\\
f_d &=& - \left(\frac{3\rho_g C_D |\dv|}{8 a_\bullet \rho_\bullet}\right)\dv.   \label{eq:drag-acelleration}
\end{eqnarray}

In the above equations $G$ is the gravitational constant, $\varSigma_g$ and $\varSigma_p$ are 
the vertically integrated gas density and bulk density of solids, respectively; 
$\v{u}$ stands for the velocity of the gas parcels; $\v{x}_p$ is the position 
and $\v{v}_p$ is the velocity of the solid particles, $P$ is the vertically
integrated pressure, $c_s$ is the sound speed, $\varPhi$ the gravitational
potential, $\nu$ the viscosity, and $\vt{S}$ the rate-of-strain tensor. The functions 
$f_D(\varSigma_g)$ and $f_\nu(\v{u},\varSigma_g)$ are sixth order hyperdiffusion and 
hyperviscosity terms that provide extra dissipation near the grid scale, explained 
in Lyra et al. (2008a). They are needed because the high order scheme of 
the Pencil Code has too little overall numerical dissipation. 

The function $\v{f}_d$ is the drag force by which gas and solids interact.
In \Eq{eq:drag-acelleration}, $\rho_\bullet$ is the internal density of a 
solid particle, $a_\bullet$ its radius, and $\dv=\v{v}_p-\v{u}$ its 
velocity relative to the gas. $C_D$ is a dimensionless coefficient that 
defines the strength of the drag force. We use the formula of Woitke \& 
Helling (2003) that interpolates between Epstein and Stokes drag

\begin{equation}
  C_D = \frac{9\Kn^2 C_D^\Eps + C_D^\Stk}{(3\Kn+1)^2}.
  \label{eq:coeff-general}
\end{equation}where $C_D^\Eps$ and $C_D^\Stk$ are the coefficients of Epstein and Stokes drag, respectively. These are
 
\begin{eqnarray}
  C_D^\Eps &\approx& 2\left(1+\frac{128}{9\pi \Ma^2}\right)^{1/2} \label{eq:coeff-epstein}\\
  C_D^\Stk&=&\left\{ \begin{array}{ll}
    24\,\Rey^{-1} + 3.6\,\Rey^{-0.313}  & \mbox{; $\Rey \leq 500$};\\
    \xtimes{9.5}{-5}\,\Rey^{1.397} & \mbox{; $500 < \Rey \leq 1500$};\\
    2.61  & \mbox{; $\Rey > 1500$}; \end{array} \right. \label{eq:coeff-stokes}
\end{eqnarray}where $\Ma=|\dv|/c_s$ is the Mach number, 
$\Rey= 2a_\bullet\rho_g |\dv|/\mu$ is the Reynolds number of the flow past 
the particle, and $\mu=\sqrt{8/\pi}\rho_g
c_s \lambda/3$ is the kinematic viscosity of the gas. We defer the reader to 
Lyra et al. (2008c) for further details of the drag force. The self-gravity 
solver is also explained in that paper.

\subsection{Initial and boundary conditions}
\label{sect:initial-conditions}

In this paper, we use the Pencil Code{\footnote{See http://www.nordita.org/software/pencil-code}} 
in Cartesian and cylindrical geometry. The cylindrical runs do not include the gravity of 
the particles and were therefore 
only used for tests or runs without particles, as will become clear in the next sections. 
A Cartesian box was used for the production runs. The Cartesian box ranges 
$x,y\in[-2.0,2.0]r_0$. The resolution is 256$\times$256, unless stated otherwise. The 
cylindrical grid ranges $r\in[0.3,2.0]r_0$, with 2$\pi$ coverage in azimuth.

The density profile follows the power law $\varSigma_g$=$\varSigma_0 r^{-0.5}$ and 
the sound speed is also set as a power law $c_s=c_{s_0} r^{-0.5}$. These slopes 
are chosen to cancel a radial dependency on the mass accretion rate 
$\dot{m}=3\pi\nu\varSigma_g$ if the viscosity follows the Shakura-Sunyaev 
parametrization, $\nu$=$\alpha c_s^2/\varOmega$. The velocity field 
is set by the condition of centrifugal equilibrium

\begin{equation}
  \dot\phi^2 = \varOmega_{\rm K}^2 + \frac{1}{r} \left[ \frac{1}{\varSigma_g}\frac{\partial{P}}{\partial{r}} + \frac{\partial{\varPhi_{\rm sg}}}{\partial{r}}     \right] 
  \label{eq:centrifugal}
\end{equation}

We use units such that $r_0$=$\varSigma_0$=$GM_\odot$=1. We choose
$c_{s_0}=0.05$ and $r_0$=5.2 AU. We impose that the disk should 
have 30 \mearthp of solid material within the modeled range. We 
also use a surface density of solids following the same slope of the 
surface density $\varSigma_p$=$\epsilon\varSigma_{0} r^{-0.5}$, where 
$\epsilon$ is the initial solids-to-gas ratio. For the interstellar ratio 
$\epsilon$=$\ttimes{-2}$, the surface density at $r_0$ is 
$\varSigma_{0}$=277 g\,cm$^{\rm -2}$ (which we round to 
$\varSigma_{0}$=300 g\,cm$^{\rm -2}$). The gas disk thus has 
a mass of $\ttimes{-2} M_\Sun$ within the modeled range. We further 
discuss the disk mass in Sect.~\ref{sect:diskmass}. The gas has  
a Toomre $Q$ parameter of 30 at $r_0$, and is throughout 
stable against gravitational instability.

The dead zone is modeled as static viscosity jumps following arc-tangent profiles 

\begin{equation}
  \nu = \nu_0 - \frac{\nu_0}{2}\left[\tanh\left(\frac{r-r_1}{\Delta{r}}\right) - \tanh\left(\frac{r-r_2}{\Delta{r}}\right)\right]
  \label{eq:viscosity}
\end{equation}where $r_1$=0.6 and $r_2$=1.2 are the locations of the jumps 
and $\Delta{r}$ its width. We adopt $\Delta{r}$=$\ttimes{-2}$, which provides 
a smooth jump over two grid cells only. The jump is thus close to a 
Heaviside function yet still differentiable.  We use $\nu_0$=$\xtimes{2.5}{-5}$ 
in code units, corresponding to an alpha value (Shakura \& Sunyaev 1973) of 
$\alpha$$\equiv$$\nu\varOmega/c_s^2$$\approx$$\ttimes{-2}$.

For the solids, we use $\ttimes{5}$ or $\xtimes{4}{5}$ Lagrangian numerical particles. 
For a gas mass of $\ttimes{-2} M_\Sun$ and the interstellar solids-to-gas 
ratio of $\ttimes{-2}$, each numerical particle therefore is a super-particle 
containing (in the lower resolution case) $\ttimes{-9} M_\Sun \simeq \xtimes{2.7}{-2} M_{\rm Moon}$ of material.  
We use particles of radii $a_\bullet$=1, 10, 30, and 100\,cm, as also used in 
Lyra et al. (2008c). For our nebula parameters, maximum drift occurs for particles 
of 30\,cm, as detailed in that paper. 

The particles are initialized as to yield a surface density following the same
power law as the gas density, and their velocities are initialized to the
Keplerian value. We use reflective boundaries for the cylindrical grid and frozen 
boundaries for the Cartesian grid (Lyra et al. 2008a). Both use the buffer zone 
described in de Val-Borro et al. (2006) to damp waves before they reach the 
boundary. Particles are removed from the simulation if they
cross the inner boundary.

\section{Vortices}
\label{sect:vortices}

\begin{figure}
  \begin{center}
    \includegraphics[width=\hfwidthsingle]{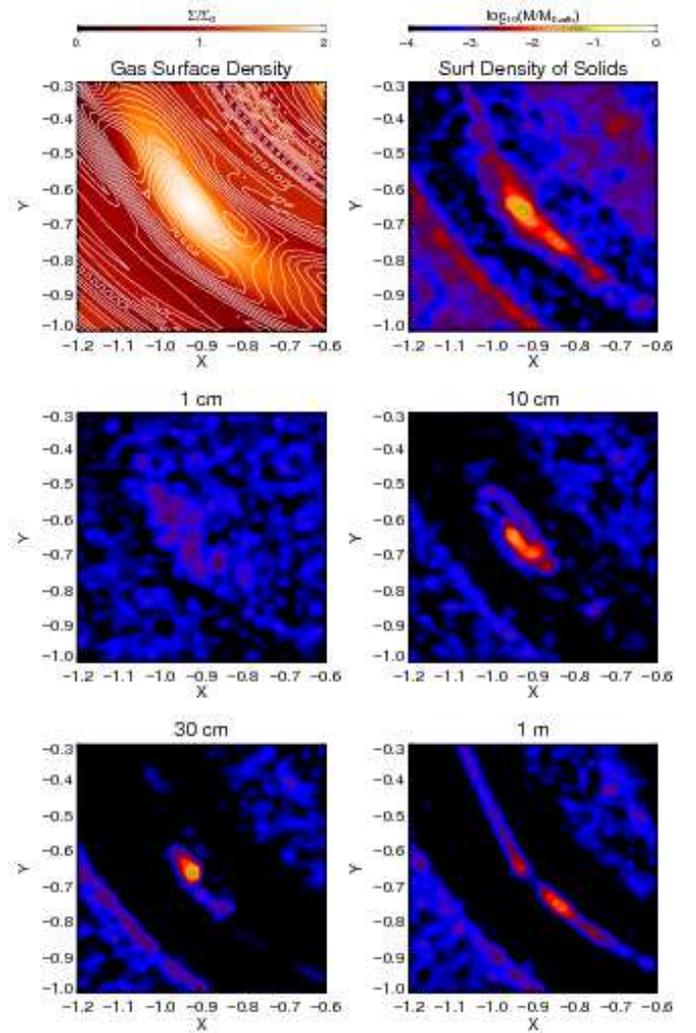}
  \end{center}
  \caption[]{Enlargement around one of the vortices in a snapshot at 20 
orbits. Contours of $|\v{u}|$ are superimposed 
on the gas surface density plot, 
showing that the density enhancement is associated with intense vorticity.  In 
the upper panel we show the multiphase (total) surface density of solids, 
whereas in the middle and lower panels we show the contribution of each 
particle species. The vortical motion preferentially traps particles of 
$a_\bullet$=10 and 30 cm.} 
  \label{fig:vortex-lines}
\end{figure}

The trapping mechanism of vortices is not only due to 
its being a high-pressure region, but mainly due to the 
vorticity of the flow. In an anticyclonic vortex (cyclonic 
vortices are destroyed by the Keplerian shear), the motion occurs in the 
same sense as the local shear, i.e., the gas rotates clockwise. Therefore, 
at the antistellar point the angular momentum is decreased with respect 
to a non-vortical flow; and conversely increased at the substellar point. 
As a result, the gas at the antistellar point is accelerated 
inwards, while the gas at the substellar point is accelerated 
outwards. A net centripetal force 
towards the eye ensues. The streamlines of vortices (or vortex 
lines) are a set of Keplerian ellipses with the same semimajor 
axis but different eccentricities, being circular in the center and 
more eccentric outwards (Barge and Sommeria 1995). We show contours 
of $|\v{u}|$ on the surroundings of one of the giant vortices, in the 
upper left panel of 
\fig{fig:vortex-lines}. As the gas drags the particles, the particles 
also revolve around the vortex eye. But because the gas-solids 
coupling is not perfect, the particles lose angular momentum and sink 
deeply towards the center. In the next subsections we describe some of the 
properties of the vortices present in our simulations

\subsection{Launching Mechanism - the RWI}
\label{sect:rwi}

The vortices in LJKP08 are triggered by 
the Rossby wave instability (RWI), a case of 
purely hydrodynamical instability in accretion disks. Considering 
azimuthal perturbations to the inviscid Euler equations, 
Lovelace et al. (1999) and Li et al. (2000) find that instabilities 
exist when the following quantity has a local extremum 
 
\begin{equation}
  \mathcal{L}(r)\equiv\mathcal{F}(r) \left(P\varSigma^{-\gamma}\right)^{2/\gamma}
\end{equation}

The quantity $\mathcal{F}$ is defined as
\begin{equation}
  \mathcal{F}\equiv\frac{\varSigma\varOmega}{\kappa^2 - \Delta{\omega}^2 - c_s^2/\left(L_s L_p\right)}
\end{equation}where 

\begin{equation}
  \kappa\equiv\left[\frac{1}{r^3}\frac{d }{dr}\left(r^4\varOmega^2\right)\right]^{1/2}
\end{equation}is the epicyclic frequency and

\begin{eqnarray}
  L_s&\equiv&\gamma\bigg/\left[\frac{d}{dr}\ln{\left(P\varSigma^{-\gamma}\right)} \right]\\
  L_p&\equiv&\gamma\bigg/\left[\frac{d}{dr}\ln{P}\right]
\end{eqnarray}are the radial length scale of the entropy and density variations, respectively. 
$\gamma$ is the adiabatic index. For corotational modes 
($\Delta{\omega}\equiv\omega-m\varOmega \ll \kappa$) in a barotropic ($L_s\rightarrow\infty$) disk, 
the quantity $\mathcal{F}$ reduces to $\varSigma\varOmega\kappa^{-2}$, which is readily identified 
with (half) the inverse of vortensity $\xi$

\begin{eqnarray}
\xi&=&\omega_z/\varSigma \\ 
\omega_z&=&|\curl{\v{u}}|_z \\\nonumber
        &=&\frac{1}{r}\frac{\partial{\white{r}}}{\partial{r}}\left(r^2\varOmega\right)=\frac{\kappa^2}{2\varOmega},
\end{eqnarray}which in turn led Lovelace et al. (1999) to interpret $\mathcal{L}$ as an 
entropy-modified version of, or generalized, potential vorticity. An extremum 
in the profile of $\mathcal L$ can be generated, for example, by a pressure bump
somewhere in the disk. The dispersion relation of the 
disturbances is analogous to the dispersion relation of Rossby waves 
in planetary atmospheres, hence the name of the instability. 

\subsection{How sharp need the jump be?}
\label{sect:sharpness}

\begin{figure}
  \begin{center}
    \includegraphics[width=\hfwidthsingle]{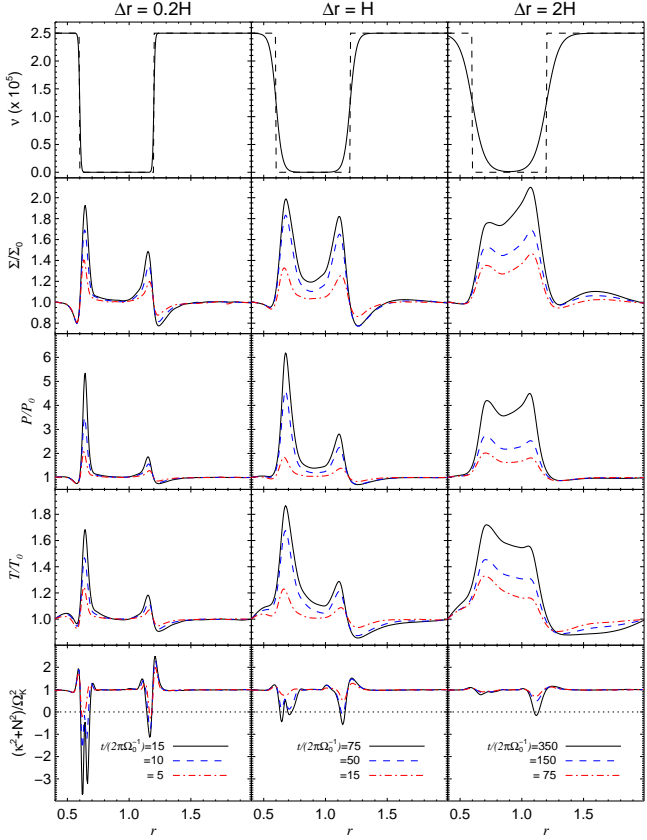}
  \end{center}
  \caption[]{Evolution of the pressure bump with different widths of the 
 viscosity jump (uppermost panels; the dashed line is the Heaviside jump, for 
 comparison). The violation of the Solberg-H{\o}iland 
 criterion (lowermost panels) is a 
 conservative indication that the threshold of instability of the RWI was 
 reached. As the width of the jump increases, the threshold takes increasingly 
 longer time to be breached. For jumps smoother than two scale lengths, the threshold  
 was not reached (up to 500 orbits).}
  \label{fig:width}
\end{figure}

Although formally the instability is triggered by a minimum or 
maximum of $\mathcal L$, ones finds in practice that the amplitude and 
radial width of the pressure bump present critical values beyond which 
instability does not occur. Li et al. (2000) find that typically, 
a pressure variation of 10-20\% over a length similar 
to the scale height of the disk is sufficient to trigger the instability. 
The threshold, however, is problem dependent, depending - among other things -
 on the geometry of the pressure variation (step jump or Gaussian, 
for example; see Li et al. (2000) for details). Due to this, we 
use a more general criterion to assess the threshold of instability. 
Li et al. (2000) note that the threshold of instability for the RWI is 
always reached before the Solberg-H{\o}iland criterion for stability of 
axis-symmetric disturbances 

\begin{equation}
  \kappa^2+N^2\geq 0
  \label{eq:sh}
\end{equation}is violated. \Eq{eq:sh}, therefore, provides a {\it conservative} estimate of whether or not the RWI 
is excited. The Solberg-H{\o}iland criterion is easily understood. In a pressureless disk, the condition $\kappa^2\geq 0$ suffices to determine 
stability. The other term 

\begin{equation}
  N^2\equiv\frac{1}{\varSigma}\frac{dP}{dr}\left(\frac{1}{\varSigma}\frac{d\varSigma}{dr} - \frac{1}{\gamma P}\frac{dP}{dr}\right)
\end{equation}is the square of the Brunt-V\"ais\"al\"a frequency, 
associated with the oscillations of buoyant structures in the presence 
of an entropy gradient. Physically, \Eq{eq:sh} means that a mode that 
is unstable/stable to shear can be stabilized/destabilized by pressure 
gradients and vice-versa.

We measure the epicyclic and Brunt-V\"ais\"al\"a frequencies in a series of 1D 
simulations where we varied the width $\Delta{r}$ of the viscosity jump 
(\Eq{eq:viscosity}). We find that for locally isothermal simulations, the 
Solberg-H{\o}iland criterion is violated at the outer edge 
for $\Delta{r} \leq 0.04$, which is slightly less than one scale 
height. In these isothermal simulations, the criterion depends 
almost solely on the epicyclic frequency because, as the temperature 
does not rise with compression, the pressure does 
not change enough for $N^2$ to go appreciably negative. 

To assess the effect of non-barotropic behavior, we replace the locally 
isothermal flow by an isentropic one with an adiabatic equation of state

\begin{eqnarray}
  \frac{\partial{S}}{\partial{t}}&=&-\left(\v{u}\cdot\del\right) S + f_\chi(S) \label{eq:entropy}\\
  P&=&\varSigma c_s^2/\gamma \label{eq:pressure}
\end{eqnarray}which means that we allow heating and cooling by compression 
and rarefaction only, excluding viscous heating and radiative cooling. 
In \Eq{eq:entropy}, $S=\ln{P} - \gamma\ln\varSigma$ is the 
vertically integrated specific entropy of the gas. The function $f_\chi(S)$ 
is a sixth order hyperconductivity term, analogous to hyperdiffusion for density. 
The adiabatic index is $\gamma$=7/5.

The results are illustrated in \fig{fig:width}, where we plot the viscosity profile 
(upper panel) for different widths $\Delta{r}$, and the time evolution of the 
density, pressure and temperature bumps. The lower panels measure if the 
Solberg-H{\o}iland criterion was violated. We find that the pressure bump 
sharpens considerably compared to the isothermal case, due to the high temperatures associated 
with the compression. The consequence is that the Solberg-H{\o}iland criterion 
is violated by viscosity jumps up to $\Delta{r} \leq 0.12$, i.e., 3 times broader than 
in the isothermal simulations. In this non-isothermal case, it is mostly the 
Brunt-V\"ais\"al\"a frequency that leads $\kappa^2 + N^2$ to negative values. The 
effect of increasing the width is mainly of slowing the evolution of the quantity 
$\kappa^2 + N^2$ towards negative values. It only takes five orbits for 
$\Delta{r}=0.01$, but it takes 350 orbits when $\Delta{r}$ is increased to 0.1. In 
\fig{fig:width} we state $\Delta{r}$ in terms of the scale height $H$=0.05\,$r_0$. We 
present a resolution study of vortex excitation in \App{app:resolution}. We also 
address the issue of vortex survival in a non-static dead zone in \App{app:vary-nu-time}. 

\subsection{Steady-state dead zone}

If no transport happens in the dead zone, matter can do little more 
than piling up there as the inflow proceeds from the active layers.  
However, the accumulation of matter cannot proceed indefinitely since, 
as matter piles up, the conditions for gravitational instability would 
eventually be met (Armitage et al. 2001). The gravitational turbulence 
that ensues (Lodato 2008) would therefore empty the dead zone as the excess 
matter accretes, thus re-starting the cycle. 

However, local simulations show that the dead zone has some level of 
residual turbulence. This happens because the turbulence on the active 
layers induce small levels of Reynolds stress in the dead zone 
(Fleming \& Stone 2003). If the inertia of the midplane layer is 
not too high (Oishi et al. 2007), this forced turbulence 
can lead to moderate $\alpha$ values with non-negligible transport {\footnote{Another 
alternative is local ionization provided by the decay of the short-lived 
radioactive nuclide $^{\rm 26}$Al (Umebayashi \& Nakano 2009), although Turner 
\& Sano (2008) show that the free electrons given out by this low ionization 
source would quickly recombine on the surface of $\mu$m-sized dust grains.}}.

Terquem (2008) shows that steady state solutions in 1D models exist in this 
case, as the dead zone gets denser and hotter to match the condition 
of constancy of the mass accretion rate with radius, $\partial_r{\left(\nu\varSigma\right)}$=0. In this case, the steady state will have an $\nu_T$ viscosity 
value in the active layers and a lower $\nu_D$ in the dead zone.

Vortex formation by the RWI requires the presence of a pressure maximum.
In our model, and that of Varni\'ere \& Tagger (2006) and Inaba \&
Barge (2006), the pressure maximum comes about by stalling the
accretion flow in the border of the dead zone. There is no requirement
that the dead zone should have zero viscosity, just a viscosity
significantly lower than that of the active regions. We tested
different values of $\nu_D$/$\nu_T$, and
found that changing it from 0 to 0.1 has little effect on the
instability. For higher values, the Solberg-H{\o}iland criterion takes
increasingly longer to be violated.  For $\nu_D/\nu_T=0.5$, the
Solberg-H{\o}iland criterion is violated after 60 orbits. We also notice 
that the steady-state dead zones of Terquem (2008; see Fig.~3 of that paper)
have the surface density varying by more a factor of $\sim$10 over a few
scale lengths at the inner edge. Such profiles violate the
Solberg-H{\o}iland criterion, so the RWI is expected to be excited in
those conditions as well. 

\section{Embryos}
\label{sect:embryos}

\subsection{Drag force cooling and compactness}
\label{sect:compactness}

\begin{figure}
  \begin{center}
    \includegraphics[width=\hfwidthsingle]{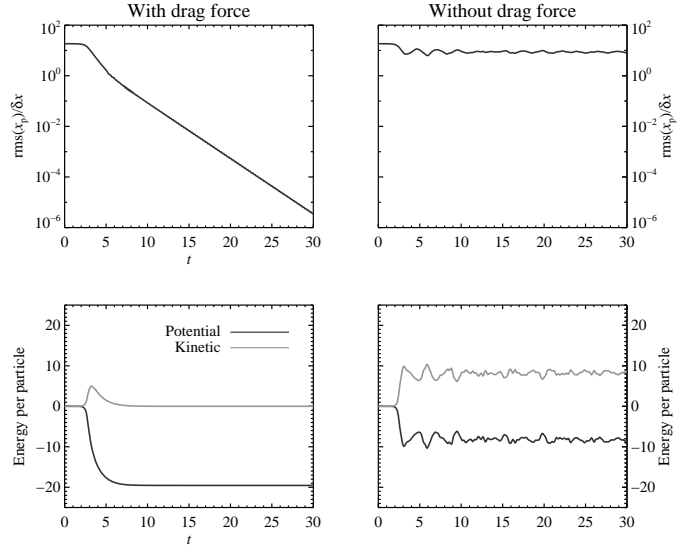}
  \end{center}
  \caption[]{1D simulation of collapse of 1000 particles. With drag force
the kinetic energy of the particles is efficiently dissipated and the 
particles collapse at subgrid scale towards infinite density. When the 
drag force is excluded the system cannot dissipate energy and a halo of 
particles, 10 grid cells wide, is formed.}
  \label{fig:collapse}
\end{figure}

\begin{figure}
  \begin{center}
    \includegraphics[width=\hfwidthsingle]{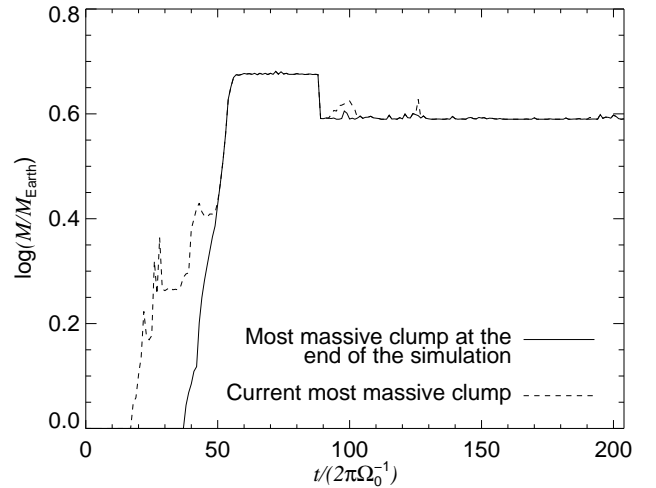}
  \end{center}
  \caption[]{Evolution of the most massive clump formed (solid line), which 
we traced back in time from the end of the simulation. It differs from the instantaneous most massive clump (dashed line) because the clumps have different feeding rate and can also experience mass loss, as in the episode that happened at $\approx$ 90 orbits (see text).}
  \label{fig:follow-clump-mass}
\end{figure}

The embryos formed in our simulations present a number of interesting features. 
We first would like to address the issue regarding their physical size. The 
embryos consist of a cluster of a large number of particles, held together by 
their collective gravitational pull. But are they strongly bound like solid 
objects? Or do they consist of loosely coupled objects in the 
same potential well? To answer this question we measure the rms spatial 
dispersion of the particles inside the cluster, defined as 

\begin{equation}
  r_{\rm rms}= \sum_{i=1}^{n} |\v{r}_i-\v{r}_{_{\rm CM}}|
  \label{eq:rms-radius}
\end{equation}where $n$ is the number of particles within the Hill sphere of 
the clump, $\v{r}$ is the vector radius of each particle and $\v{r}_{_{\rm CM}}$ is the vector 
radius of their center of mass. We take this value as a 
measurement of the ``radius'' of the embryo. The most massive embryo has 
a radius of 1.13 $R_\oplus$. This compactness corresponds to a tenth of a thousandth 
of the resolution element.

Such compactness is due to the dynamical cooling provided by the drag force. 
We illustrate this in \fig{fig:collapse}. The figure shows the results of 
1D simulation with a thousand interacting particles with and without gas 
drag. Without gas drag the particles have no means to dissipate energy and 
perform oscillations about the center of mass. The very inner 
particles show virialization, while the outer particles form a halo 
extending for a radius of 10 grid cells in average. 

When including gas drag, the system gets so dissipative that the kinetic 
energy is soon lost and the ensemble of particles collapses. The exponential 
decay of the 
particles' rms position seen in the upper left panel of \fig{fig:collapse} shows 
no sign of flattening, down to a millionth of a resolution element. This 
leads us to infer that collapse to zero volume is ongoing. This is of 
course expected, since no mechanism to provide support against the gravitational 
pull is present. 

In view of this, the question is why our planets, which are 
subject to drag forces, do not shrink to zero as well but stabilize at a 
very small but finite radius. We are drawn to the conclusion that this is 
a numerical issue. The tests of \fig{fig:collapse} were done with a fixed 
time-step. But when the particles cluster together to form a planet in 
our simulations, they end up dominating the time-step. The position 
update $\v{x}(t)=\v{x}_0 + \v{v}\Delta{t}$ therefore occurs 
with the maximum $\Delta{t}$ allowed by the Courant 
condition, which is that the fastest particle should move by one grid cell. 
Due to this, the time resolution of the subgrid motion around the center of 
mass of the cluster is under-resolved. With the overshot $\Delta{t}$, 
the particles that are attracted towards the center 
of mass of the clump will end up in a position {\it past it}. In the next time 
step they will be attracted to the center of mass from the other side, but will 
once again overshoot it. The result of this is that the particles will execute 
undamped oscillations, leading to a finite rms radius. We performed 
tests like those of \fig{fig:collapse} with a particle-controlled variable 
timestep, confirming this explanation. We conclude that the fact that the most 
massive embryo has a stable rms radius compatible with its mass is but a deceptive 
coincidence. 

We stress that this drag force cooling will cease to be efficient as 
the solids-to-gas ratio grows too large ($\rho_p/\rho_g \gg 1$), because in 
this case the backreaction would be too strong and the gas would simply be 
dragged along with the particles. In this case, a Keplerian disk of solids
might form, accreting matter onto the planet due to collisions between the
orbiting solids. This accretion regime is nevertheless beyond the scope of the
current paper.

\subsection{Mass loss}
\label{sect:mass-loss}

In Fig.~2a of LJKP08 we showed the evolution of the most massive clump. 
However, as the clumps have different feeding rate and some of them experience 
mass loss, the most massive clump shown there is not always the same clump. 
In \fig{fig:follow-clump-mass} we contrast this with the evolution of the 
most massive clump {\it at the end of the simulation}, which we tracked backwards 
in time. Such clump started in the inner disk, showing 0.8 $M_{\rm Mars}$ by 40 orbits. 
By this time, the most massive clump was a 3 $M_{\rm Mars}$ clump in the outer disk. 

\begin{figure}
  \begin{center}
    \includegraphics[width=\hfwidthsingle]{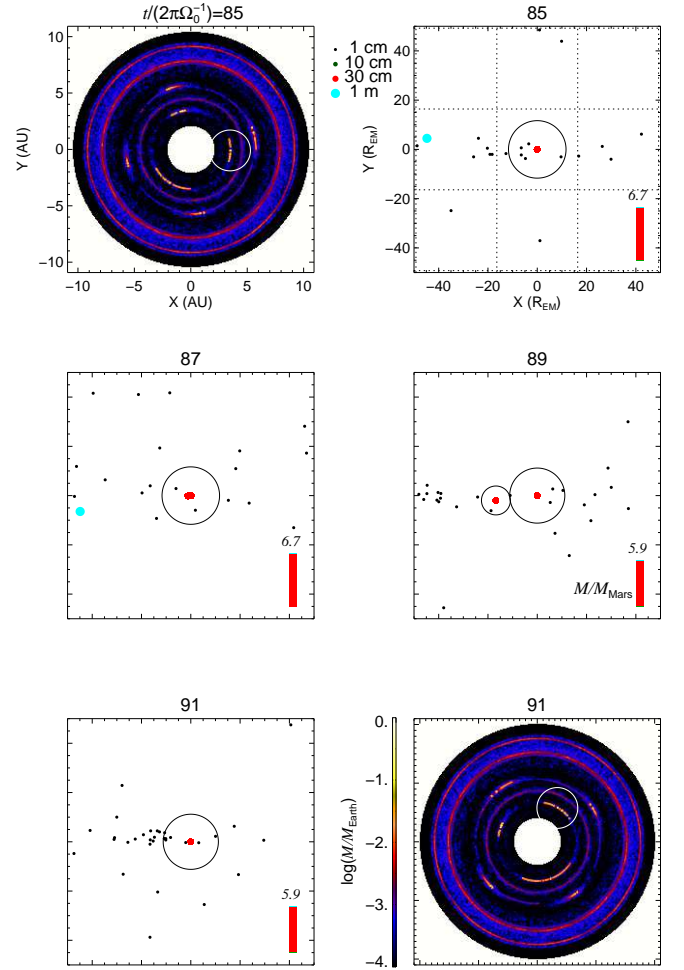}
  \end{center}
  \caption[]{Time series of the mass loss episode. Due to gravitational tides from the 
massive vortex whence 
the embryo formed, a large chunk of particles was detached from the original cluster. 
At 87 orbits, a separation 4 times the Earth-Moon distance is seen. The separation grows 
and two orbits later the two bodies do not overlap Hill radii, thus counting as separate 
objects.} 
  \label{fig:massloss}
\end{figure}

The most remarkable feature of this plot is the mass loss event at 90 orbits. 
\fig{fig:massloss} shows that it consists of the detachment of a 0.8 $M_{\rm Mars}$ 
object from the original cluster, of 6.7$M_{\rm Mars}$. The detachment is already 
seen at 87 orbits, although the separation is quite small (4 times the Earth-Moon mean 
separation, $R_{\rm EM}$). At 89 orbits, 
the smaller object left the Hill sphere of the more massive embryo. They finally appeared as 
separate objects, and the maximum mass decreased.

\begin{figure*}
  \begin{center}
    \includegraphics[width=.8\textwidth]{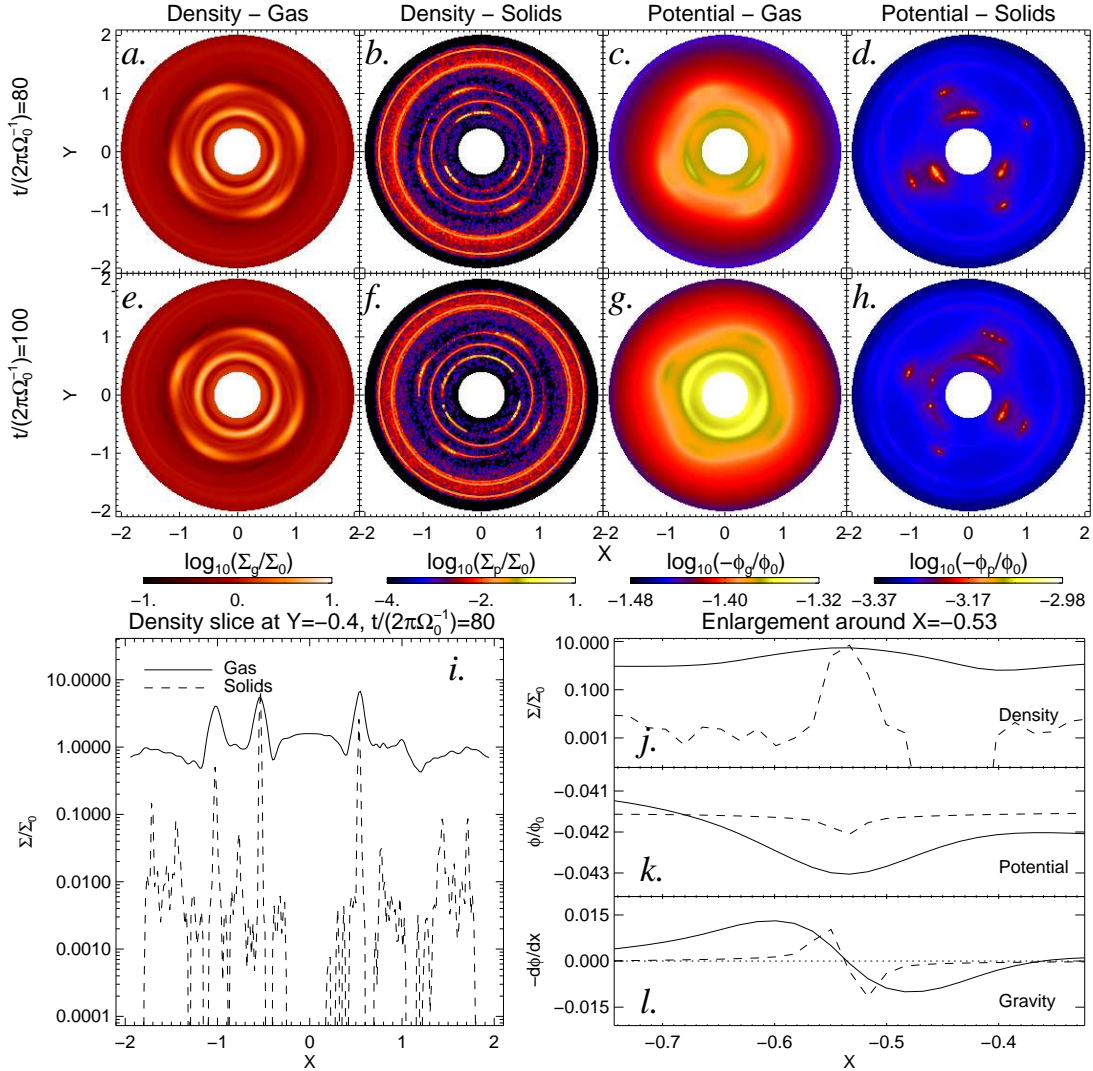}
  \end{center}
  \caption[]{The state of the disk before ($a$-$d$) and after ($e$-$h$) 
the mass loss episode. The conspicuous difference between them 
is that the embryo has left its parental vortex from one snapshot to the 
other. It is seen as a bright spot in panels $f$ and $h$, 
at (X,Y)=(-0.65,-0.19). In panel $b$ (before the mass loss), the embryo 
is at (X,Y)=(-0.40,-0.53) but not easily spotted among the swarm of solids 
inside the vortex. Panel $i$ shows a horizontal slice through this 
location, in which we see that the density of solids does not peak much 
higher than the gas density at the location of the embryo (panel $j$). 
Significant gas tides are expected as the gravitational potential 
(panel $k$) and acceleration (panel $l$) have similar contributions 
from the gas and solid components.}
  \label{fig:tides}
\end{figure*}

We see evidence that this puzzling behavior is due to gravitational tides from the gas. 
The gas is too pressure-supported to undergo collapse, but the vortices concentrate 
enough material to yield a considerable gravitational pull. We illustrate this 
in \fig{fig:tides}, where we show the state of the disk before the mass loss episode 
(at 80 orbits, \fig{fig:tides}a-\fig{fig:tides}d) and after that (at 100 orbits, 
\fig{fig:tides}e-\fig{fig:tides}h). 
The plots show the surface densities of gas and solids, and the potential associated 
with them. Even though the clumping of solids yield a considerable gravitational 
pull (\fig{fig:tides}d and \fig{fig:tides}h), these figures show that the dominant 
contribution to the gravitational potential comes from the gas - more specifically 
from the vortices, where the gas density peaks one order of magnitude denser than the 
initial condition. 

The most massive clump is located in the inner disk at 
(X,Y)=(-0.40,-0.53) in \fig{fig:tides}b, not clearly identifiable 
amidst the other particles trapped inside the vortex. However, the embryo is 
immediately observable as the bright point at (X,Y)=(-0.65,-0.19) in \fig{fig:tides}h 
(also visible in \fig{fig:tides}f, albeit less prominently). 
There are two features in this plot that are worth noting. 
First, by comparing the location of the embryo in these figures with the location of the 
vortices, we notice that 
the planet has left its parental vortex. Second, the inner 
vortices have undergone the transition from the $m$=3 to the $m$=2 mode. Due to 
merging, their gas density has increased, with dramatic consequences for the 
embryos within them. 

We assess how the gravity of the gas influences the motion of the particles 
(\fig{fig:tides}i-\fig{fig:tides}l). In \fig{fig:tides}i we take a horizontal 
density slice at the position 
of the most massive embryo at 80 orbits. \fig{fig:tides}j is an enlargement of 
\fig{fig:tides}i around X=-0.53, where the embryo is located. We see that the 
densities of solids and gas peak at similar values. The next figures show the gravitational 
potential (\fig{fig:tides}k) and acceleration (\fig{fig:tides}l) around the embryo. 
The gas produces a deeper gravitational well, albeit smoother than the one displayed 
by the solids. In the acceleration plot it is seen that the pull of the gas is greater 
than the pull of the embryo already at a distance of just 0.26 AU (0.03 in code units, 
corresponding to two grid cells) away from the center. And even where the pull of 
the solids is strongest (one grid cell away from the center), the gravity of 
the gas still is an appreciable fraction of the gravity of the solids. Tides from 
the gas are unavoidable. 

It is beyond the scope of this paper to consider the full 
mathematical details of the theory of tides, especially because the two bodies (the vortex and the 
embryo) are extended. Instead, we consider the following toy model. The tidal force 
$F_T$ experienced by the planet is proportional to the gradient of the acceleration $a$
induced by the gas. It is also proportional to the radius $R$ of the planet: 
$F_T \propto R\,\grad{a}$. Since $\grad{a}=-\Laplace{\varPhi}$, according to 
the Poisson equation, the tidal force is proportional to the local value of the density

\begin{equation}
  F_T \propto R\,\rho_g . 
\end{equation}

We consider the 3D volume density to avoid the requirement of using the Dirac 
delta in the 2D case. Considering 
the planet spherical, Newton's second theorem holds and we can write $F_G= -GM/R^2$
for the planet's (self-)gravitational force at its surface. Substituting $M =4/3 \pi \rho_p R^3$, 
we have $F_G\propto R \rho_p$, so 

\begin{equation}
  \zeta=\frac{F_T}{F_G} \propto \frac{\rho_g}{\rho_p},
  \label{tide-to-selfgrav}
\end{equation}i.e., the ratio of the disrupting 
tidal stresses to the self-gravitating forces that attempt 
to keep the planet together is directly proportional to the gas-to-solids 
ratio. At 80 orbits, as seen in \fig{fig:tides}j this ratio is around unity. 
As the vortices undergo the transition from the $m$=3 to the $m$=2 mode, their peak density 
increases (while the planet remains at a constant mass), the tides 
eventually become strong enough to cause the mass loss event of 
\fig{fig:massloss}. This effect will probably be 
less dramatic in 3D simulations because, as the particles settle in the midplane, 
the ratio of the volume gas density to the bulk density of solids 
$\rho_g$/$\rho_p$ is expected to be much lower than the ratio of the 
column gas density to the vertically integrated surface density of solids 
$\varSigma_g/\varSigma_p$.

\subsection{Erosion?}
\label{sect:erosion}

\begin{figure}
  \begin{center}
    \includegraphics[width=\hfwidthsingle]{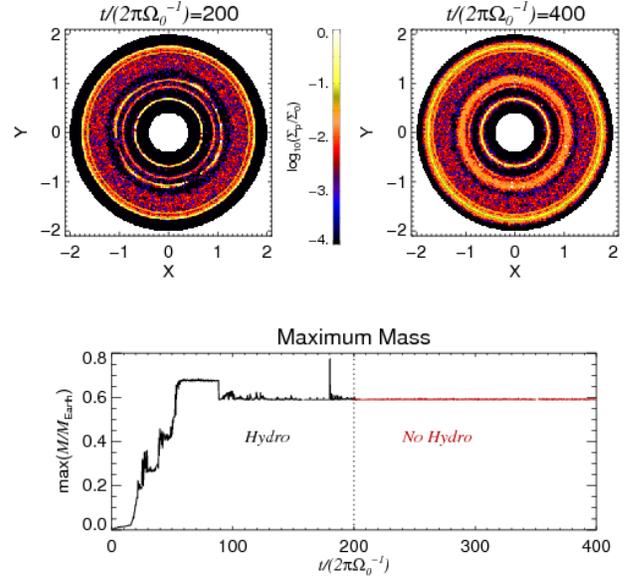}
  \end{center}
  \caption[]{Can self-gravity alone maintain the formed embryos together? 
We switch off the hydrodynamics at the last snapshot and run for additional 
200 orbits. The most massive embryo, which had already left the parental 
vortex (at $\approx$ 90 orbits), does not disperse.}
  \label{fig:shutdownhydro}
\end{figure}

Cuzzi et al. (2008) points that erosion is of prominent importance in the 
stability of self-gravitating clumps of particles. They put forth a model 
where self-gravity plays a role analogous to that of surface tension in 
liquid drops, preventing disruption against ram pressure forces from the gas. 
The clumps are held together by self-gravity if the gravitational Weber number 
(in analogy with the surface tension case) is less than a critical value, close 
to 1. The gravitational Weber number is defined as the ratio of the drag to 
self-gravitational accelerations

\begin{equation}
  {\rm We}_G = \frac{|f_d|}{|\grad{\varPhi_{\rm sg}}|}.
  \label{eq:weber}
\end{equation}

Cuzzi et al. (2008) further point that in numerical models, artificial viscosity can largely exaggerate 
the disrupting effect of the ram pressure. This happens because, as the clumps 
are small, they are deeply within the viscous range of the grid, whereas in the real 
solar nebula the dissipation happens at much smaller scales. The Reynolds number 
of the flow past the particles is therefore much smaller than what a real clump would 
experience {\footnote{The Reynolds number of 
the flow past a clump can be written as 
$\Rey=R_{\rm rms} v_{\rm rms}/\nu$. At the grid scale our choice of viscosity 
is usually $\xtimes{3}{17}$ cm\,s$^{\rm -1}$ (it decreases very fast as we go to 
smaller wavenumbers, as $k^{6}$).  For a clump as the ones of this study, of 
$R_{\rm rms}$=$\ttimes{4}$\,km and $v_{\rm rms}$=1 m\,s$^{\rm -1}$, the 
Reynolds number is $\Rey\approx\xtimes{3}{-7}$. At such incredibly low 
Reynolds numbers, inertia plays no role. The self-gravity of the particles, 
therefore, is not holding the cluster together against drag forces from 
the gas, but against largely exaggerated viscous stresses. In comparison, in the 
solar nebula, the molecular viscosity is much lower and the Reynolds 
number is expected to be $> \ttimes{6}$ (Cuzzi et al. 2008).}}, and the 
exaggerated viscous stresses might de-stabilize
potentially stable clumps.

It is interesting to assess if this mechanism plays a significant role in our 
models, or even if it can account for at least some of the mass loss events. 

We can estimate how important erosion will be for our clumps the following way. 
We approximate the clumps as single point masses so that $|\grad{\varPhi_{\rm sg}}|\approx GM/r^2$, where $M$ is 
the total mass of the clump. Plugging this in \eq{eq:weber}, we write the gravitational Weber number as

\begin{equation}
  {\rm We}_G = \frac{3\rho C_D |\Delta v|^2 r^2}{8\,G M\,a_\bullet \rho_\bullet}.
\end{equation}

For Epstein drag (\eq{eq:coeff-epstein}), $C_D$ does not depend on $a_\bullet$. So,
 for all other quantities being constant, we expect ${\rm We}_G$ to decrease 
linearly as $a_\bullet$ increases. In other words, self-gravitating clumps 
of larger particles should be more stable than clumps composed of smaller particles.
 
We can simplify the ${\rm We}_G$ by writing the mass $M$ as $M$=$\pi r_{\rm rms}^2 \varSigma_p$, 
and the drag force as $|f_d|$=$\Ma\,c_s/\tau$. Thus, at $r$=$r_{\rm rms}$,

\begin{equation}
  {\rm We}_G = \frac{\Ma\,c_s}{\tau \pi G \varSigma_p}.
  \label{eq:weber}
\end{equation}

We confirm in  
\App{app:erosion} that \eq{eq:weber} is sufficiently accurate in predicting the onset
of erosion. For our choice of parameters, 

\begin{equation}
  {\rm We}_G \approx \frac{17\,\Ma}{(\tau\varOmega)\varSigma_p\,r^2}
  \label{eq:weber2}
\end{equation}so for a flow of $\Ma\approx\ttimes{-2}$, a clump of $\varSigma_p$=1 
at $r\approx 1$ will be stable if $\tau\varOmega\apprge0.1$. The embryos we consider 
are formed predominantly of particles of 
10\,cm ($\tau\varOmega$$\approx$0.1) and 30\,cm ($\tau\varOmega$$\approx$0.3). 
We conclude that erosion might play a role in our case. The sharp dependence of ${\rm We}_G$ on the distance in \eq{eq:weber2} also means that embryos at the inner edge 
of the dead zone are more prone to erosion than the ones at the outer edge. We will 
develop this further in Sect~\ref{sect:tidal-disruption}. 

\subsection{Can the embryos be held together indefinitely?}
\label{sect:shutdown-hydro}

To answer the question of how long lived these clumps are, we 
take the last snapshot of the simulation and switch off the 
hydrodynamics. The particles now move under the influence of the stellar 
gravity and their own self-gravity only. We run for additional 200
orbits to assess is self-gravity alone can maintain the cluster of 
particles together. The result is shown in \fig{fig:shutdownhydro}.

The clustered particles do not disperse, and the most massive embryo 
maintains the same mass for the additional simulating time. We do not 
see difference in the degree of compactness of the most massive embryo. 
As the vortices are shut down, the unbound 1\,cm sized particles that were 
too well coupled to the gas to be dragged into the eye are released from they 
vortical confinement and spread over a wider annulus. 

We have no reason to suspect that the situation will change over longer 
timescales. We conclude that the embryos can be held together indefinitely.

\subsection{Collisions}
\label{sect:collisions}

\begin{figure}
  \begin{center}
    \includegraphics[width=.9\hfwidthsingle]{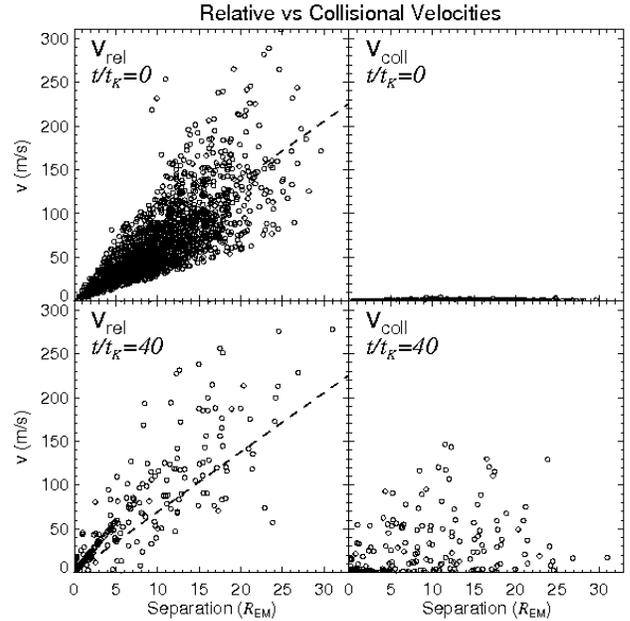}
  \end{center}
  \caption[]{The relative velocities between particles are contrasted with 
collisional velocities calculated from \Eq{eq:collision-velocity}.  At the 
initial condition ($t$=0) the relative velocities between particles follow 
what is expected from the Keplerian shear (dashed line), reaching as far 
as 300 \mps inside a grid cell (which is 31 $R_{\rm EM}$ wide). In contrast, 
the true collisional velocities are much lower. The figure shows that 
estimating fragmentation with the relative velocities would greatly 
overestimate its likelihood. Time is quoted in orbits ($t_K$=$2\pi/\varOmega_0$).} 
  \label{fig:velcoll}
\end{figure}

As stated before, one of the problems that solids accumulation inside vortices
can potentially solve is the issue of fragmentation of particles upon
collisions, a drawback for both coagulation (Brauer et al. 2008a) and
gravitational instability models (Rice et al.  2006, Johansen et al.
2007) of planetesimal growth.

To assess if fragmentation poses a significant barrier for the 
formation of the protoplanetary embryos in this study, we take the 
most massive embryo by the end of the simulation, flag the particles 
that constitute it and trace them back in time, calculating their 
collisional velocity history. The collisional speed for each particle 
is calculated by taking the closest neighbor to that particle 
within the range of a grid cell. 
There is, however, a subtlety concerning the difference between collisional 
velocities and relative velocities. A collision 
between particles $i$ and $j$ only happens when the separation $r_{ij}$ 
between them tends to zero. For our resolution and choice of $r_0$, a 
grid cell is 0.08 AU wide, thus existing plenty of room for subgrid 
Lagrangian dynamics. In particular, the velocity difference due to 
the Keplerian shear between the inner and outer radial borders of a 
grid cell can be significant. At the inner edge of the dead zone 
of the model presented in this paper (3.12 AU) for instance, this 
difference amounts to 434 \mps. As this velocity difference is due 
to the separation between particles, it 
vanishes when $r_{ij}$ tends to zero, thus never contributing to 
the true collisional speed. In \fig{fig:velcoll} we show these uncorrected 
relative velocities in the initial condition and in a snapshot at 40 orbits, 
plotted against separation. The clear correlation follows what is expected 
from the Keplerian shear (dashed line) in the initial condition. 

\begin{figure*}
  \begin{center}
    \includegraphics[width=\textwidth]{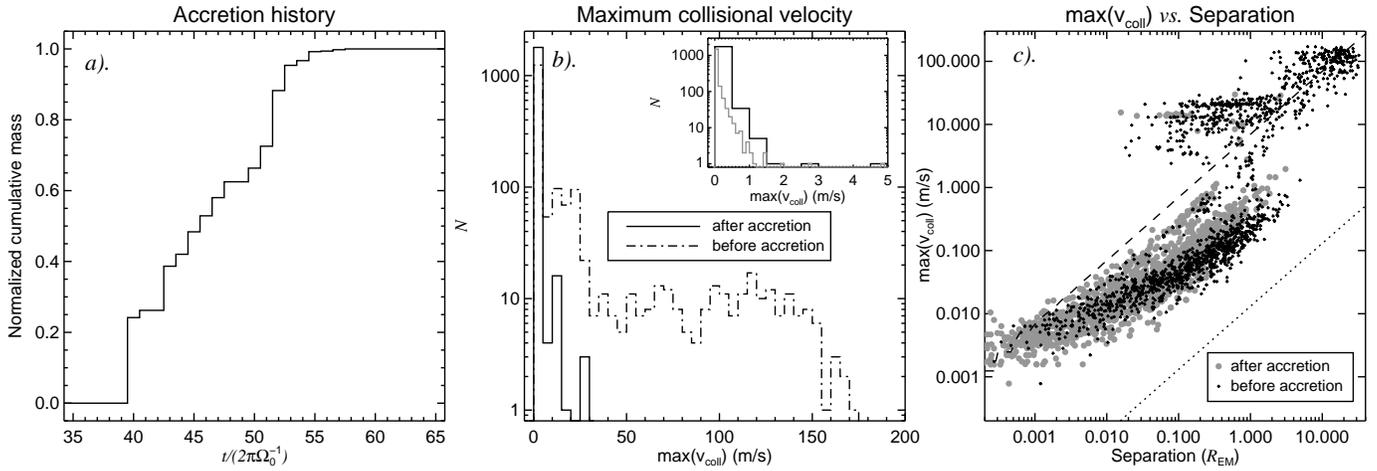}
  \end{center}
  \caption[]{{\it a).} Accretion history of the most massive embryo. 
Collapse happens at 40 orbits, and further accretion of particles 
happens through the next 18 orbits, after which the maximum mass 
is attained. This plot defines the time $t_0$ that each particle is accreted.\\ 
    {\it b).} Histograms of the maximum collisional velocities 
experienced by a given particle, before (dot-dashed line) and after (solid line) 
its accretion time $t_0$. The solid line histogram represents the maximum collisional 
speed from $t_0$ until the end of the simulation. The dot-dashed line histogram 
covers a time interval of ten orbits before $t_0$. As the vortices form at 
$\approx$30 orbits, the latter better represents $v_{\rm coll}$ under vortex trapping. The vast 
majority of the particles are in the bin of 0-5 \mps. The smaller window zooms 
into this bin. The black line corresponds to bins of 1\mps whereas the gray line 
to bins of 0.1\mps. Both represent $v_{\rm coll}$ after $t_0$. It shows that 
most of the particles never experience collisions more violent than 1 \mps.\\
{\it c).} Same as {\it b).} but plotting the maximum collisional speeds as function 
of separation between the particle and its nearest neighbor. Before $t_0$, 
three populations are seen, of low speed at short separations, high 
speeds at short separations, and high speeds at large separations. Only the 
second group would have experienced destructive collisions. After $t_0$, 
99\% of the particles belong to the first group. The correlation with distance for the first 
group is not due to the Keplerian shear (dashed line) or the residual shear present 
in $\Delta{v}$ (dotted line).}
  \label{fig:collision}
\end{figure*}

\begin{figure}
  \begin{center}
    \includegraphics[width=\hfwidthsingle]{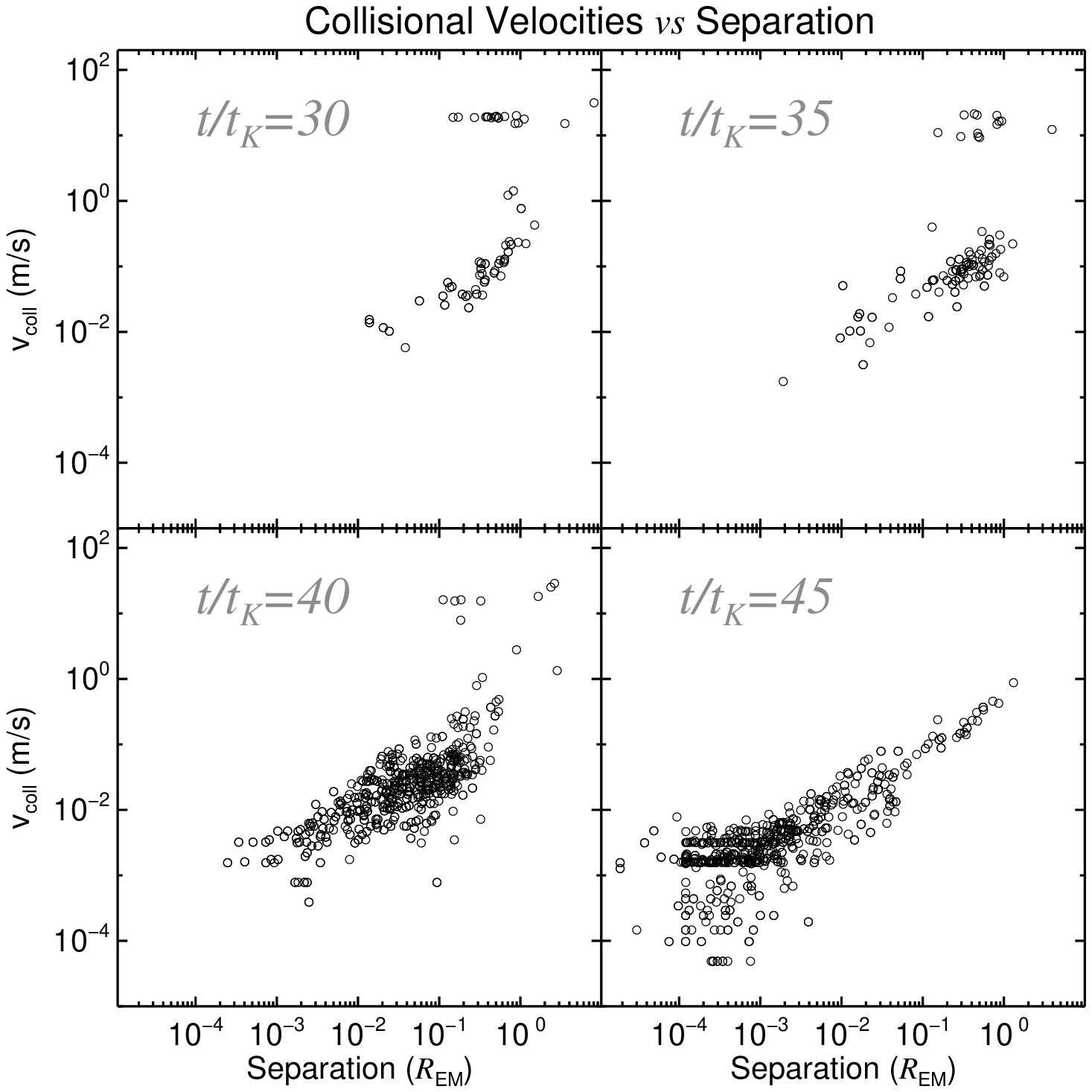}
  \end{center}
  \caption[]{Collisional speeds as a function of separation at selected snapshots at 
30, 35, 40, and 45 orbits ($t_K$=$2\pi/\varOmega_0$). The figure only shows particles 
that are closer than the distance of a grid cell to the clump of maximum number 
density. The group at low separations and high speeds observed in the upper plots 
would undergo fragmentation. Nevertheless, they represent only a minority of 
particles (22\% at 30 orbits, 8\% at 35). At the time the cluster gets bound 
(at 40 orbits), the vast majority of particles (99\%) is safely at low speeds. The 
presence of a fourth group at 45 orbits, at very low separations ($<$$\ttimes{-3}\,R_{\rm EM}$)
and with very low collisional speeds ($<$$\ttimes{-3}$\,\mps) indicates that collapse 
towards zero volume is ongoing.} 
  \label{fig:collision2}
\end{figure}

The gas motion adds another velocity that has to be taken into 
account. Solid particles are dragged by the gas motion, yet gas 
streamlines never intersect. The gas motion itself 
thus introduce velocities that never participate in collisions. We 
correct for these two by the following procedure. For each particle 
in the pair involved in a collision, we consider its velocity 
$\Delta{\v{v}}$ relative to the gas (the same quantity that appears in the 
drag force, $\Delta{\v{v}}$=$\v{v}_p-\v{u}$). We then define the 
collisional velocity vector as 

\begin{equation}
\v{v}^{ij}_{\rm coll} \equiv \Delta{\v{v}}(\v{x}_{p_i}) - \Delta{\v{v}}(\v{x}_{p_j}).
\label{eq:collision-velocity}
\end{equation}

Equation~(\ref{eq:collision-velocity}) ensures that 
tracer particles, which follow the gas 
streamlines, should never experience collision. Furthermore, as the shear 
dependence of $\Delta{\v{v}}$=$\eta\v{v_K}$ is much smaller than the 
shear dependence of $\v{v}_p$=$\v{v_K}$ (where 
$\eta$ = $(1/2)(H/r)^2(\partial\,{\ln P}/\partial\,{\ln r})$ = $\xtimes{3.75}{-3}$ 
is the pressure-correction factor), we can assume that it also corrects for 
this quantity. Figure~\ref{fig:velcoll} also shows the collisional speeds 
calculated by \Eq{eq:collision-velocity}. The dependence on separation 
was greatly suppressed. 

The results of the collisional velocity history of the particles 
that constitute the most massive embryo at the end of simulation 
are plotted in \Fig{fig:collision}. 

\Figure{fig:collision}a shows a cumulative plot of the mass 
of the embryo, which defines the time $t_0$ that each flagged particle 
was accreted. We define the time of accretion as the moment when the 
particle approached the grid point $\v{x}_{\rm near}$ nearest to the maximum 
of particle number density (also defined by the flagged particles) by 
less than $d_{\rm diag}$=d$x\sqrt{2}$ (the grid cell diagonal) and 
kept 
\begin{equation}
|\v{x}_p - \v{x}_{\rm near}| \leq d_{\rm diag} 
\label{eq:condition}
\end{equation}until the end of the simulation. Although this is not as strict as the definition of 
accretion we have been using before (based on the Hill criterion and 
escape velocity), this simpler criterion captures what happens before 
collapse (i.e., before the maximum of particle number density becomes 
a bound protoplanetary embryo) and serves well our purpose of 
illustrating the behavior of collisional speeds at close separations.
The first episode of accretion takes place at 40 orbits, coinciding 
with the time that the clump of particles became bound (in accordance 
with \fig{fig:follow-clump-mass}). Further accretion proceeds over the next 
18 orbits, with the maximum mass being attained at 58 orbits. No other 
particle was accreted after this time. 

\Figure{fig:collision}b shows 
histograms of the maximum collisional speed that a particle experienced 
before (dot-dashed line) and after (solid line) accretion. The latter 
refers to the maximum of $v_{coll}$ taken between $t_0$ 
and the end of the simulation (200 orbits). The former refers to a 
time interval of 10 orbits before $t_0$. As the vortices in the inner 
disk just fully develop at $\approx$30 orbits, the dot-dashed histogram 
is more representative of a situation where particles are inserted in a 
disk with existent vortices. The conclusion is striking: the vast 
majority of the particles that constitute the embryo never experienced 
a collision more violent than 1 \mps.

In \fig{fig:collision}c we show the maximum collisional velocities 
of \fig{fig:collision}b as a function of the separation between a given 
particle and its closest neighbor, also before and after accretion. 
The distribution before accretion is trimodal, with 
particles with low speeds ($<$1\mps) at small separations, particles 
with high speeds ($<$20\mps) at small separations, and particles with 
high speeds at large separations ($>$10$R_{\rm EM}$). Only the second 
group of particles would have undergone fragmentation. The first group 
is below the fragmentation velocity threshold, whereas the large 
separations of the third group imply they never experienced an encounter 
close enough to lead to a collision. After accretion, virtually all particles 
(99\%) belong to the first group.

In \fig{fig:collision2} we plot the collisional velocities versus 
separation at selected snapshots instead of historical maxima. In these plots, we 
only used the particles for which \Eq{eq:condition} was satisfied, i.e., 
considering only the collisions that are participating on the formation process of the 
embryo. At 30 orbits, a small number of particles is observed (87), 78\% of these 
showing safe collisional speeds ($<$10\mps). 5 orbits later 119 particles are within 
the grid cell of the forming embryo, 92\% of which show gentle collisions. At the time 
the overdensity gets bound (40 orbits), it is formed by 639 particles, with just 7 of 
these showing collisional speeds greater than 10\mps. At 45 orbits, all 877 particles 
display safe speeds. The tendency seen in this time series towards an increasing 
number of encounters at low separations and at low collisional speeds indicates 
that collapse towards zero volume is ongoing. Indeed, at 70 orbits, we observe that 
most of the particles occupy the same point in space. 

\section{Size distribution}
\label{sect:size-distribution}

In our simulations, we considered the solid phase of the disk represented by 
particles of 1, 10, 30, and 100 cm. In this section 
we discuss a number of issues, relevant to the simulations, related to 
having a size spectrum instead of single-phasing. 

\subsection{Aerodynamical sorting}
\label{sect:sorting}

One of the most prominent features of the embryos formed in our models is 
that they are composed primarily of same-sized particles. This is mostly 
due to aerodynamical sorting.  As particles of different size have different 
friction times, differential drag occurs inside the vortex, effectively 
sorting the particles spatially by size. Moreover, the stationary point 
is determined by a balance between the Coriolis and the drag force, in 
such a way that the eye of the vortex is the stationary point only 
for $\tau_s\rightarrow 0$, or perfect coupling. In 
general, the stationary point is azimuthally shifted with respect to 
the eye, according to the particle size (Youdin 2008).

\begin{figure}
  \begin{center}
    \includegraphics[width=\hfwidthsingle]{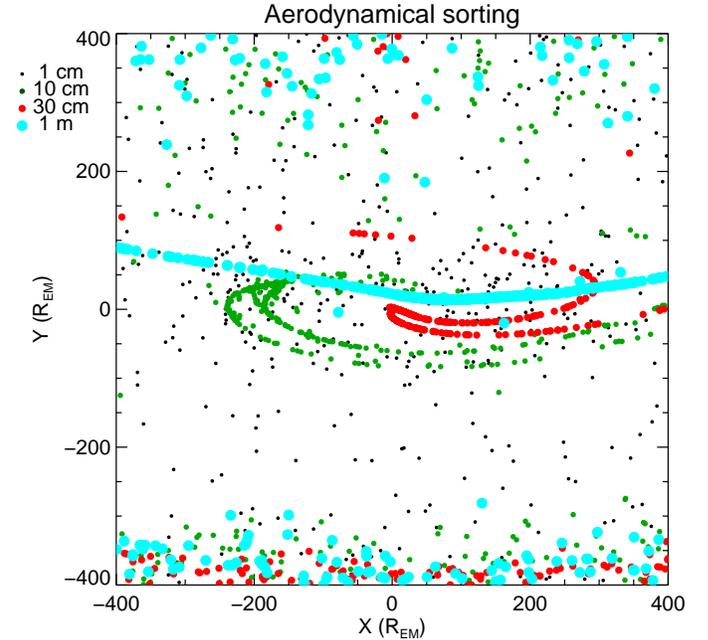}
  \end{center}
  \caption[]{Aerodynamical sorting for the particles trapped in the vortical 
motion. The figure is centered at the vortex shown in \fig{fig:vortex-lines}, 
at 18 orbits. The unit of length is the Earth-Moon mean distance, and the 
Y-coordinate points to the star. As particles of different size have different 
friction times, differential drag occurs inside the vortex, effectively 
sorting the particles spatially by size.}   
  \label{fig:aerodrift}
\end{figure}

The aerodynamical sorting inside the vortex can be seen in 
\fig{fig:aerodrift}, which corresponds to the vortex of 
\fig{fig:vortex-lines}. As similar particles drift alike, 
streams of same-sized particles are clearly seen in the vortical 
flow.

\subsection{Differences between the inner and outer embryos}
\label{sect:inner-outer}

Another interesting feature of our results is that once the vortices are formed, they easily trap 
the 10\,cm and 30\,cm particles both in the inner and in the outer edge of the dead zone. Yet, by 
the end of the simulation there seems to be a preference for the embryos in the inner disk to be 
composed of larger particles (30-100\,cm), while in the outer disk, more embryos formed of the smaller 
particles (1-10\,cm) are seen. We explain these two features in the following sections.

\subsubsection{Preferential sizes in different locations of the nebula}
\label{sect:tauomega1}

The first feature (inner and outer vortices equally trapping particles of same size) follows from the fact 
that the general drag coefficient (\Eq{eq:coeff-general}) yields a nearly flat profile for the radius of the 
particle with maximum drift ($\tau\varOmega$=1) versus distance. This is seen by calculating the stopping 
times $t_s$ of the different particles as a function of the dynamical variables

\begin{eqnarray}
  t_s&=&\tau\varOmega \nonumber\\
     &=&\frac{\sqrt{32 \pi}}{\Kn^\prime \Ma} \frac{\lambda \rho_\bullet}{\varSigma_g}
     \frac{\left(\Kn^\prime +1\right)^2}{\left(\Kn^{\prime 2} C_D^{\rm Eps} + C_D^{\rm Stk} \right)} 
\end{eqnarray}and calculating the radius for a given stopping time $t_s$. 
Substituting $C_D^{\rm Eps}$=$16/(3\Ma)\sqrt{2/\pi}$ (\Eq{eq:coeff-epstein} at subsonic regime) 
and Stokes law $C_D^{\rm Stk}$=$24/\Rey$ (with $\Rey=3\Ma/\Kn\sqrt{\pi/8}$) yields the quadratic equation 
for $a_\bullet$

\begin{equation}
  2a_\bullet^2 + 3\lambda a_\bullet - \frac{6\lambda\varSigma_g t_s}{\pi\rho_\bullet} = 0
\end{equation}and, as the radius is positive, the solution is unique 

\begin{equation}
  a_\bullet = \frac{3\lambda}{4}\left(\sqrt{1 + \frac{16\varSigma_g t_s}{3\pi\lambda\rho_\bullet}} - 1 \right)
\end{equation}
  
\begin{figure*}
  \begin{center}
    \includegraphics[width=.9\hfwidth]{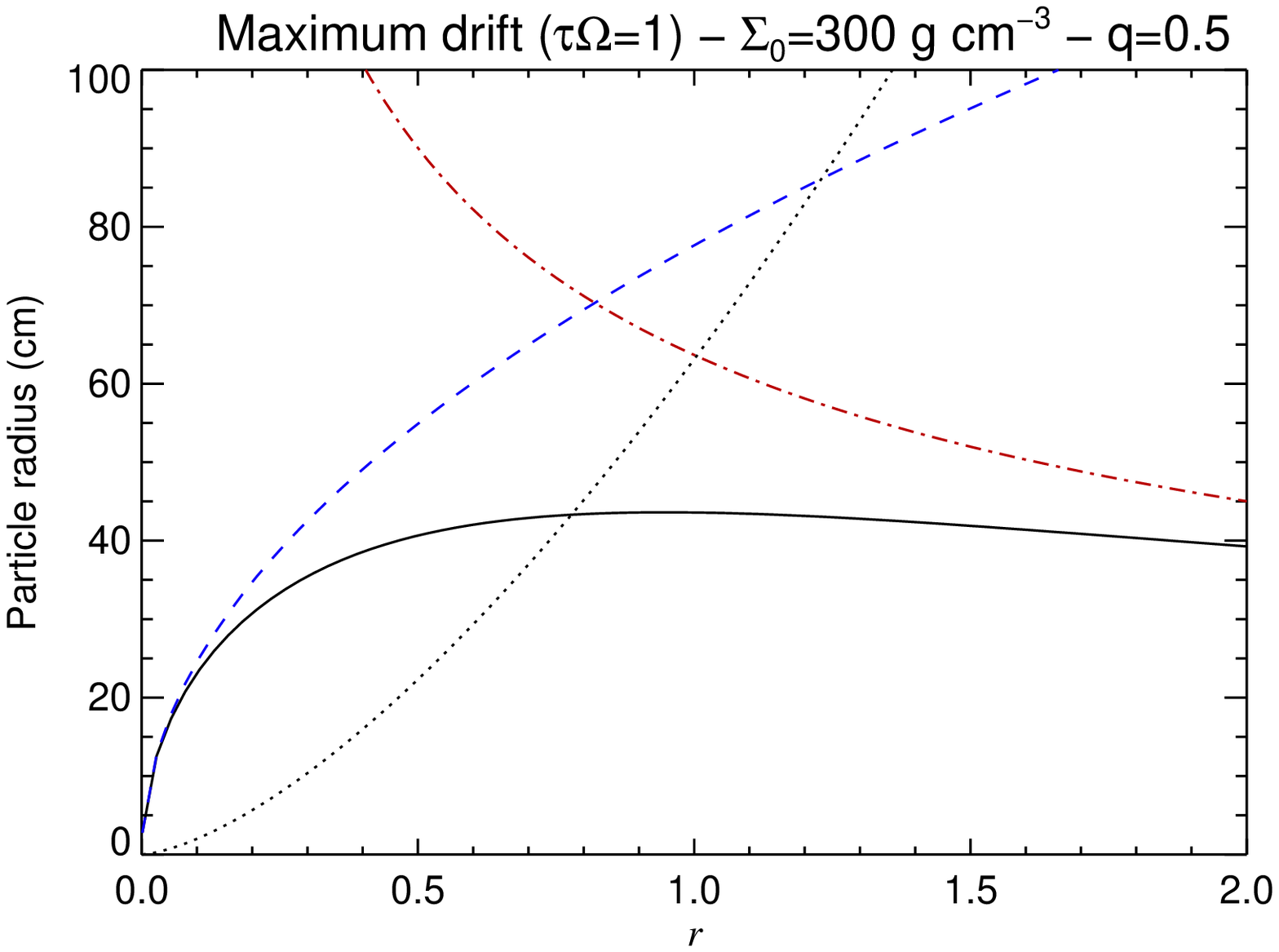}
    \includegraphics[width=.9\hfwidth]{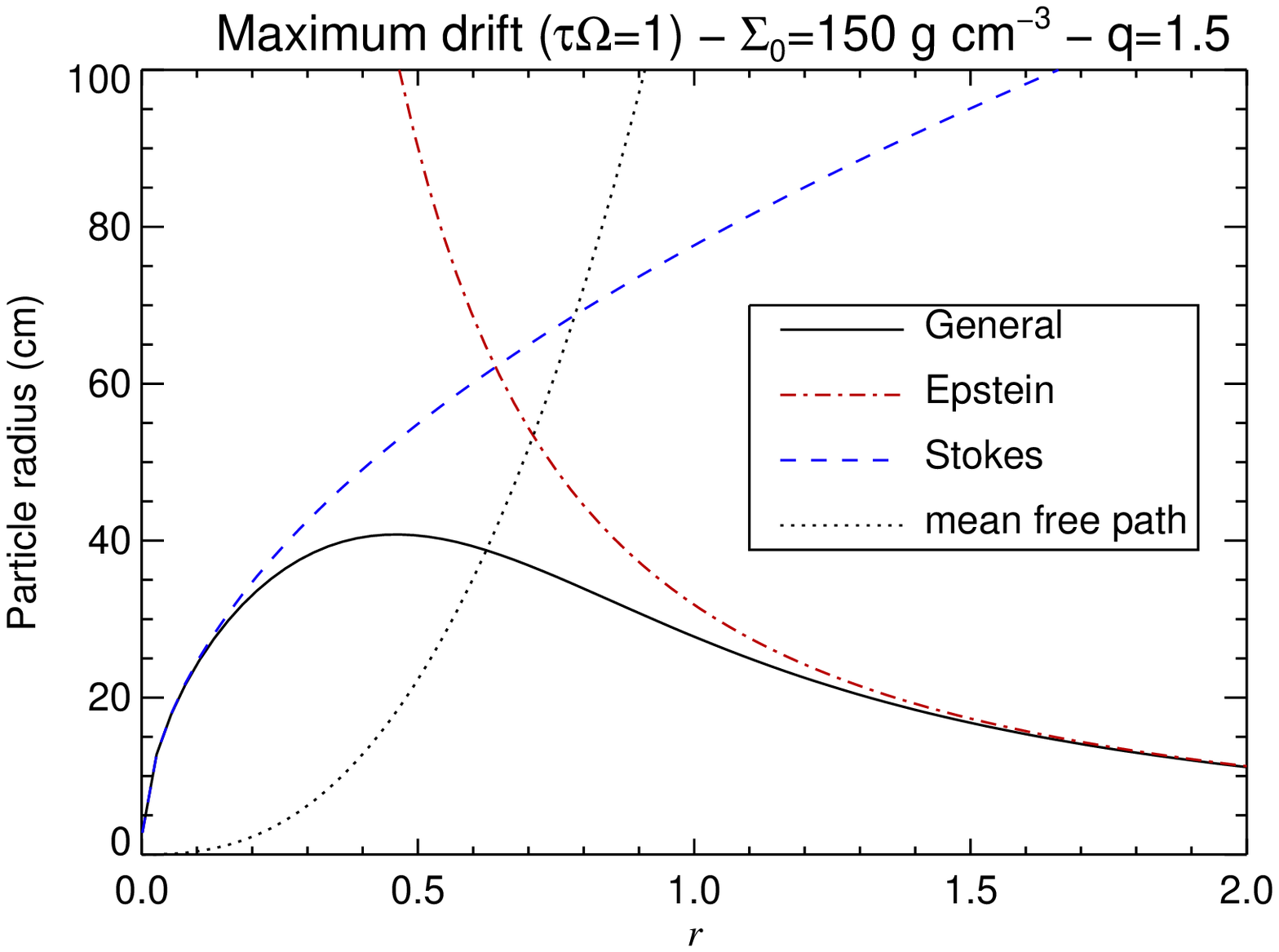}
  \end{center}
  \caption[]{The radius of the particle subject to maximum drift ($\tau\varOmega$=1) for our choice of 
parameters (left panel) and in the MMSN (right panel). $q$=$-\partial\ln\varSigma/\partial\ln{r}$ is the 
power law of the surface density profile. The profile is very flat compared to the ones predicted by the 
limiting cases of Epstein and Stokes drag, especially for our choice of parameters. The vortices 
in the inner and outer edge of the dead zone should have similar efficiency on trapping particles 
of a given size.} 
  \label{fig:preferred-rad}
\end{figure*}

\Fig{fig:preferred-rad} shows this curve as a function of radius for $t_s$=1 and our initial parameters. The 
particle radius of $t_s$=1 predicted by both pure Epstein and pure Stokes drag 
are shown for comparison. The mean free path $\lambda$ is also shown. The figure shows that the 
curve is quite flat for our choice of parameters, so inner and outer 
vortices must have similar efficiency on trapping particles of a given size.

\subsubsection{Tidal disruption and erosion of the inner embryos of $a_\bullet$=10\,cm}
\label{sect:tidal-disruption}

\begin{figure}
  \begin{center}
    \includegraphics[width=\hfwidthsingle]{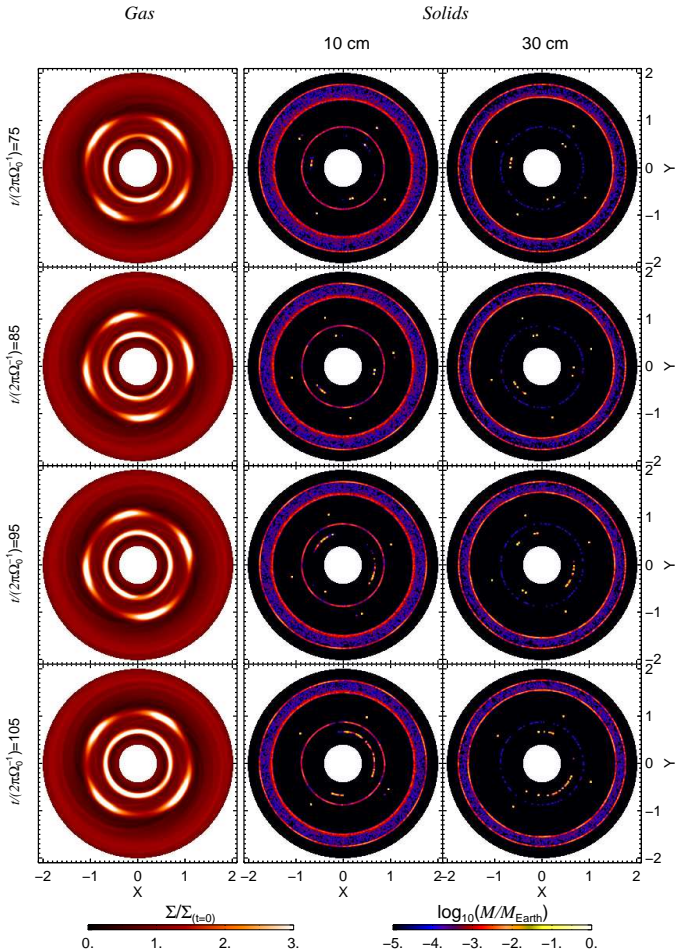}
  \end{center}
  \caption[]{Time series of the gas and solid density (for each individual particle size), showing the 
destruction of the embryos of 10\,cm particles when the inner vortices undergo the transition from 
the $m$=3 to the $m$=2 mode. At 75 orbits (upper panels), embryos are seen in both the inner and outer 
vortices, for both 10\,cm and 30\,cm particles. At 85 orbits one of the embryos in the 10\,cm phase was 
disrupted, followed by a second at 95 orbits, and the last one ten orbits later. The embryos composed of 
30\,cm particles also experience tides. But as their gravitational Weber numbers are smaller, they just 
undergo splitting, the large fragments being more stable against erosion than the embryos composed of 
particles of  10\,cm.}
  \label{fig:transition_m_3-2}
\end{figure}

We see evidence that the second feature (the absence of clumps of 10\,cm in the inner disk 
at later times even though they were formed) is due to tidal disruption in the same episode 
that lead the most massive embryo to lose mass. We illustrate this in \Fig{fig:transition_m_3-2}, 
a time series of the gas and the solid phases, the latter split into the 4 different particle 
sizes. The upper plots, at 75 orbits, show that embryos of 30\,cm and 10\,cm were formed in 
both the inner and outer vortices. 

The gas plots of the time series of \Fig{fig:transition_m_3-2} illustrate 
the transition from the $m$=3 to the $m$=2 undergone by the inner vortices, as 
mentioned in Sect. 4.2. This raises their density so the tides get stronger. 
The plots of the 30\,cm phase show that the embryos composed of these particles 
split into smaller objects, which nevertheless can still keep their physical integrity. 

The fate of the clusters composed of particles of 10\,cm is different, though. As 
the gas density increases, so does the gravitational Weber numbers of the embryos. 
Erosion starts to play a more significant role. As the density increases inside the 
inner vortices (a factor 5 relative to the initial condition at 75 orbits; 8 at 200 
orbits), the embryos of 10\,cm particles start to behave more and more like the 
$\tau\varOmega$=0.01 clusters of \fig{fig:erosion}. At 85 orbits, one of embryos 
of 10\,cm particles in the inner edge was destroyed. At 95 orbits, a second  
embryo was disrupted. At 105 orbits, the third embryo was also destroyed 
by the combined effect of tides and erosion. 25 orbits later, the 10\,cm particles 
have dispersed through the inner edge of the dead zone. The tides from the gas prevent 
them from assembling once again. 

The outer vortices never get as strong as the inner ones. The result is that although the 
inner embryos of 10\,cm particles are destroyed, the outer ones are kept until the end 
of the simulation.

\section{The response of the RWI to the drag backreaction and self-gravity}
\label{sect:rwi-response}

\begin{figure}
  \begin{center}
    \includegraphics[width=\hfwidthsingle]{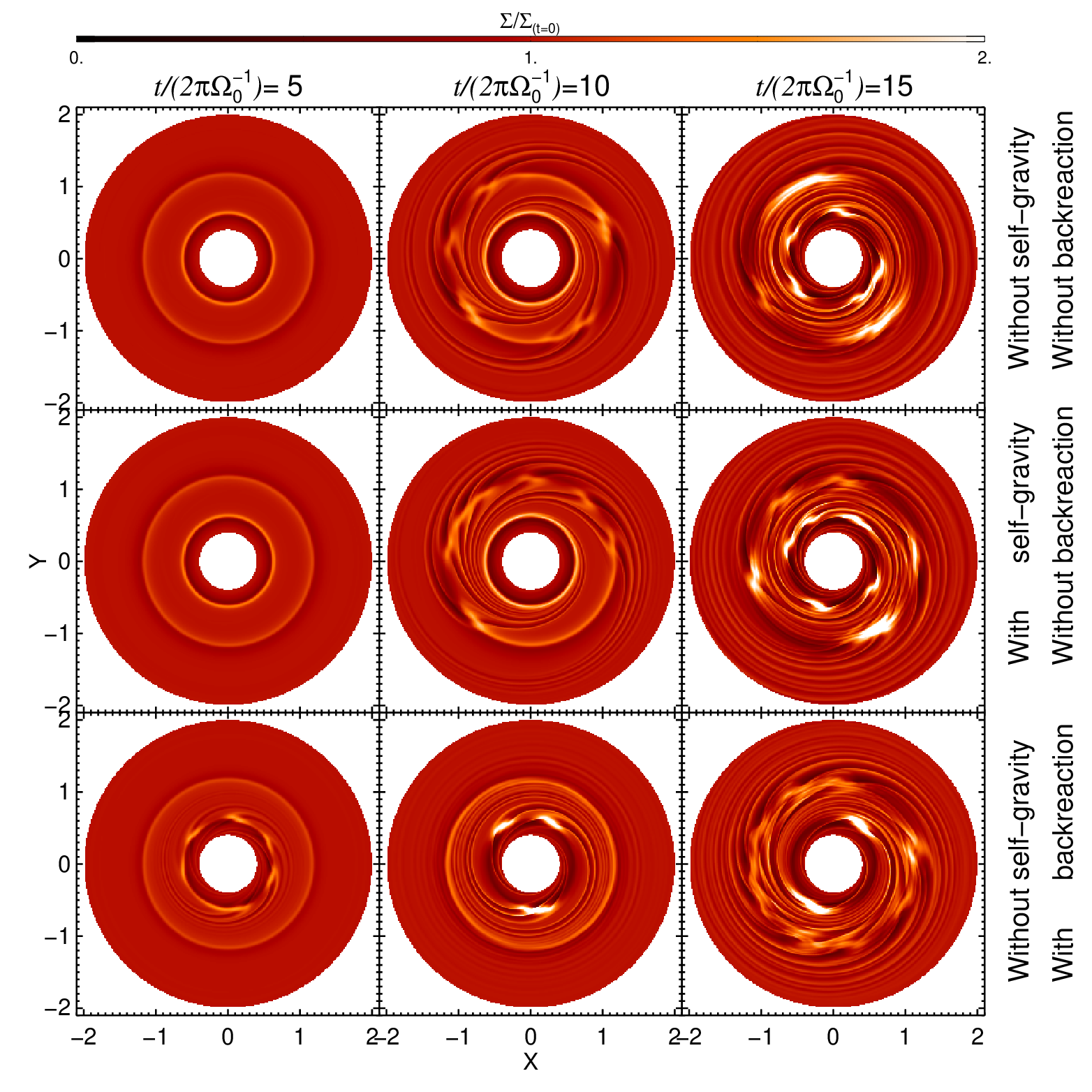}
  \end{center}
  \caption[]{{\it Upper panels}. Evolution of a disk without particles and 
without self-gravity, which serves as a control run for the next plots.\\
      {\it Middle panels}. Evolution of a disk without particles but with  
self-gravity. The difference compared to the upper panels is that the 
dominant mode in the outer disk changed from $m$=2 to $m$=5. Self-gravity 
modifies the dispersion relation of the RWI, or it stalls the inverse 
cascade of power known to occur in 2D turbulence, or both.\\
      {\it Lower panels}. Evolution of a disk without self-gravity but with  
particles. The backreaction leads to an early excitation of the RWI in 
the inner edge of the dead zone. Conversely, the outer edge goes unstable 
later when compared to the other two runs. Since the particle density is 
not high enough to excite the streaming instability, we take it as evidence 
that the backreaction modifies the dispersion relation of the RWI.}
  \label{fig:RWI-self-breact}
\end{figure}

The evolution of the RWI was studied analytically for the case of a 
low mass dustless disk only. In \fig{fig:RWI-self-breact} we show 
how the effects of gas self-gravity and backreaction from the solids 
affect the evolution of the instability. 

The upper panels of \fig{fig:RWI-self-breact} show a disk 
without solids and without self-gravity. In the middle panels 
we included self-gravity, while in the lower panels we included 
solids. The appearance of the disk in the three simulations is 
shown in selected snapshots at 5, 10, and 15 orbits. 

The self-gravitating and non-selfgravitating dustless cases (upper and 
middle panels) look similar, with the RWI being excited first in the 
outer edge of the dead zone. However, there is a crucial difference between 
them. The snapshot at 15 orbits shows a prominent $m$=$2$ mode in 
the outer disk for the non-selfgravitating case, while the run with 
self-gravity displays a dominant $m$=$5$ mode at the same time. This 
puzzling result is made even more interesting by recalling that
the gas is gravitationally stable ($Q$$\approx$30). That the growth rates of different modes 
vary that significantly for such a value of $Q$ is indicative that the dispersion relation 
of the RWI is probably remarkably sensitive to self-gravity. 

The simulation with drag backreaction (lower panels) also displays a 
number of interesting features. First, the RWI was excited in the 
inner disk as early as 5 orbits. In contrast, the control run 
without solids (upper panels) has the instability appearing first in the 
outer disk, and at later times (10 orbits). The conclusion is that the 
particles induce vorticity on their own. Even though it is clear that this behavior 
has to do with free energy being transfered from the particle motion to the 
gas motion, it is not obvious if this result can be linked to the 
streaming instability (Youdin \& Goodman 2005) since the solids-to-gas 
ratio is not nearly as high as the one needed to excite it 
($\rho_p/\rho_g \apprge 1$). Instead, it is more likely that the backreaction 
is modifying the dispersion relation of the RWI. 

Another interesting feature of this run is that although the RWI was excited 
in the inner edge of the dead zone as early as 5 orbits, the outer edge just 
went unstable as late as 15 orbits. In contrast, the control run (upper panels) 
shows the 
outer disk going unstable at 10 orbits. As the backreaction 
hastens the growth of the RWI in the inner edge, it is unclear why it should 
stall it in the outer edge. One possibility is that the 
Rossby waves launched by the edge that first goes unstable interferes destructively 
with the perturbations fighting to grow in the other edge. 

The dominant mode also changed from the dusty to the dustless case. 
The latter has $m$=4 and $m$=5 modes being dominant 
in the inner edge. In the dusty case a $m$=2 mode is seen instead. 
However, since in the dustless case it is the outer edge that displays a 
$m$=2 mode, another explanation comes to mind. As the models are 2D, we 
are probably witnessing the inverse cascade phenomenon due to enstrophy 
conservation. The vortices are simply cascading energy towards the 
larger scales, so the edge that goes unstable first (outer in the 
dustless case, inner in the dusty) will also reach a dominant 
$m$=2 mode first (possibly also $m$=1 at later times). 

\begin{figure*}
  \begin{center}
    \includegraphics[width=\hfwidth]{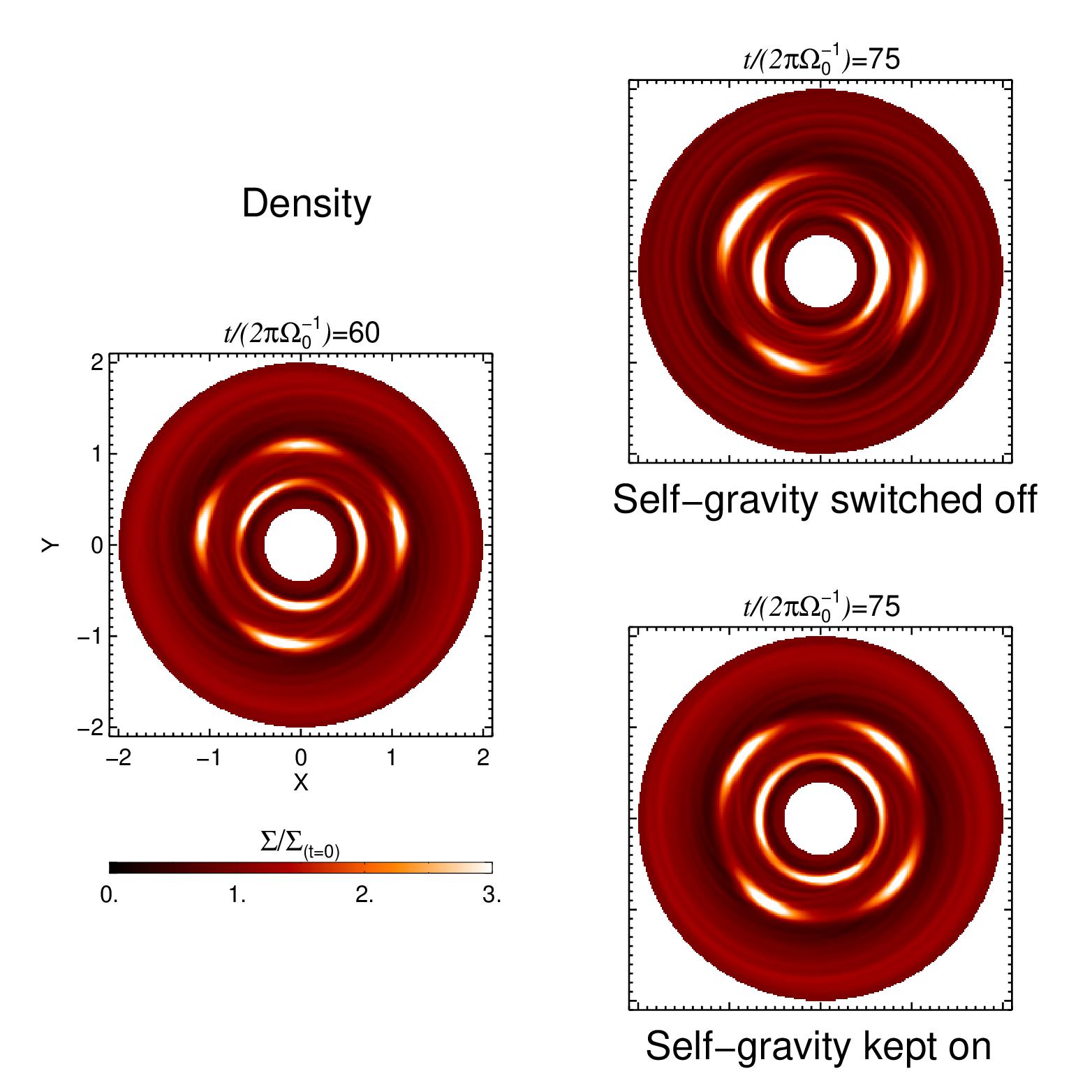}
    \includegraphics[width=\hfwidth]{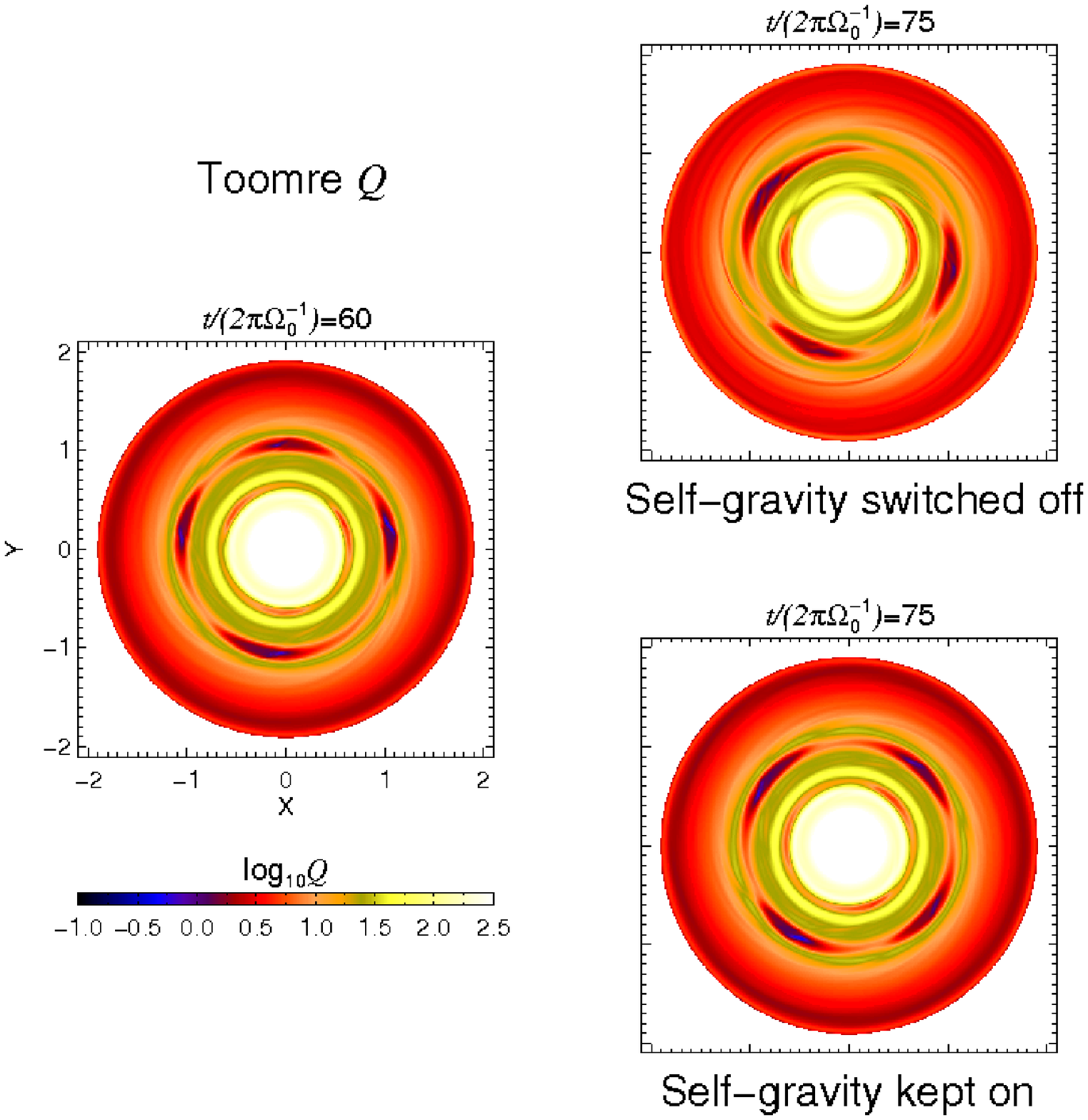}
  \end{center}
  \caption[]{In the simulation shown in LJKP08, the 
vortices maintained a $m$=4 mode in the outer edge of the 
dead zone until the end of the simulation. By switching self-gravity off, we 
see that in less than 15 orbits the outer vortices merged into a $m$=3 mode, 
while the $m$=3 mode in the inner disk edge into 
a $m$=2 mode. This is because as $Q$ decreases, 
the size of the vortices approaches the Jeans length scale, which effectively 
halts the inverse cascade of energy.} 
  \label{fig:cascade}
\end{figure*}

If this is the case, then self-gravity somehow halts the inverse cascade 
that took place in the dustless non-selfgravitating case. It is also instructive 
to compare the dusty non-selfgravitating run (lower panel) with the dusty 
selfgravitating run of LJKP08 (Fig.~1 of that paper). In that case, the $m$=4 was dominant 
in the outer edge of the dead zone until the end of the simulation at 200 orbits. 
We also perform a test (\fig{fig:cascade}) that consists of switching the 
self-gravity off in the run of LJKP08, and checking the evolution of 
the vortices. Without self-gravity, the $m$=4 mode turns into a $m$=3 mode in 
less than 15 orbits. In the inner edge, a $m$=2 mode developed out of the otherwise 
dominant $m$=3 mode. The inverse cascade indeed resumed. 

This result was also very recently reported by Mamatsashvili \& Rice 
(2009). Without self-gravity, the vortex size is limited by the pressure 
scale height $H$. Once vortices grow to sizes of a few times $H$, the vortical 
flow becomes super-sonic. The vortex then radiates density waves that carry 
energy away and limit further growth. Mamatsashvili \& Rice 
(2009) point that in the presence of self-gravity, the Jeans length
$\lambda_J \sim QH$, where $Q=(\kappa c_s)/(G\pi\varSigma)$ is the 
Toomre $Q$ parameter, poses another limitation to the maximum 
size of a vortex. In \fig{fig:cascade} we see that the vortices in 
the self-gravitating runs show $Q\approx 1$. The growth seen when 
self-gravity is switched off is a result of this constrain being lifted.

\section{Limitations of the model}
\label{sect:limitations}

The presented models are admittedly simplified. In this section, we state 
what we consider the main limitations of our calculations to be.

\subsection{Two-dimensionality}
\label{sect:2d}

The most stringent limitation of the models is the 2D approximation, which
leads to a number of features, stated below. 

\subsubsection{Vortex formation and survival}
\label{sect:rwi3d}

The question of the excitation and sustainability of vortices in three 
dimensions is the matter of an old, yet unsettled, debate. Once excited, 
anticyclonic vortices are easily maintained in 2D simulations where, unless 
viscosity is present, they cannot decay and will instead merge, growing in 
size in a cascade of energy towards the largest scale of box (e.g., Johnson \& 
Gammie 2005). However, three-dimensional studies in the context of 
protoplanetary disks found that tall vortex columns are destroyed, both 
in non-stratified (Shen et al. 2006) and in stratified (Barranco \& Marcus 
2005) local boxes. This phenomenon is understood as a result of the 
elliptic instability (Crow 1970, Gledzel et al. 1975, Kerswell 2002), 
by which the stretching term $(\v{\omega}\cdot\del)\v{u}$, absent in 2D, 
breaks down elliptical streamlines such as vortical flow. For a vortex 
to grow in 3D, the baroclinic term $\rho^{-2}\grad{\rho}\times\grad{p}$ has 
to counter the stretching term.

An indication that vortices can be sustained in three 
dimensions is present in the study of Edgar \& Quillen (2008). These authors 
simulate a stratified disk in spherical coordinates with an embedded giant planet. In their 
inviscid run, the RWI is excited, leading to Rossby vortices at the edges of the gap, 
much like as in the 2D runs of de Val-Borro et al (2007). The vortices 
launched in three dimensions are long lived and vertically extended, 
apparently following the same scale height as the surrounding disk. 
We remark that the MRI-generated vortex of Fromang \& Nelson (2005) is also 
seen to be long-lived in a 
unstratified global disk. The studies of Edgar \& Quillen (2008) and Fromang 
\& Nelson (2005) both use a locally isothermal equation of state, which has 
large-scale non-zero baroclinity due to the static radial temperature gradient.
Furthermore, the existence of the RWI in 3D is demonstrated by the simulations 
of M\'eheut et al. (2008).

\subsubsection{Strength of the vortices}
\label{sect:strength}

A major impact of the 2D approximation is the inverse cascade due to enstrophy conservation 
that overpowers the vortices. An in depth study of the formation, development and 
structure of Rossby vortices in 3D global accretion disks  
is needed to realistically address the issue of planet formation inside these structures. 

\subsubsection{Particle sedimentation}
\label{sect:sedimentation}

Another limitation posed by the two-dimensionally is that the particles and the 
gas have the same infinitely thin scale height. The result of this is that
the back-reaction of the drag force from the particles onto the gas is 
largely underestimated in 
our models. In 3D disks, the midplane particle layer is far denser 
due to sedimentation, so the ratio $\rho_p/\rho_g$ is far greater than the 
ratio $\varSigma_p/\varSigma_g$ used in \eq{eq:Navier-Stokes}. The stronger 
backreaction that ensues is known to excite the 
streaming instabilities if $\rho_p/\rho_g \apprge 1$ 
(Youdin \& Goodman 2005, Youdin \& Johansen 2007, 
Johansen \& Youdin 2007). This instability enhances particle clumping, 
thus aiding collapse (Johansen et al. 2007). However, the effect of this 
strong backreaction on the vortices is poorly known. 

\subsubsection{Different particle scale heights}
\label{sect:scale-height}

As the particles sediment, what sets the particle scale height is the equilibrium 
between turbulent diffusion and vertical gravity. Controlled by the drag force, 
the turbulent diffusion depends on the particle radius, and so does the equilibrium 
scale height of the solids (Dubrulle et al. 1995). Because of this, particles of 
1\,m radius settle in a thinner layer than those of 1\,cm particles. Inside a vortex, turbulent 
motions are expected to be weaker (Klahr \& Bodenheimer 2006), bringing the layer of solids 
closer to a 2D configuration, but a dependence on radius is still expected. We could not model 
this effect on our simulations.

\subsubsection{Gas tides and mass loss}
\label{sect:gas-tides-mass-loss}

The strength of the disrupting gas tides is yet another effect related 
to the difference between 2D and 3D models. As discussed in Sect.~3.2, the 
tides are proportional to the gas-to-solids 
ratio $\rho_g/\rho_p$, thus expected to be much weaker in 3D where sedimentation 
considerably increases $\rho_p$ in the midplane. As the vorticity is also expected 
to be weaker in 3D, the peak of $\rho_g$ at the vortex's eye would be weaker than 
in a 2D calculation, further weakening the effect of tides. We therefore anticipate 
embryos formed in 3D simulations to be significantly less prone to mass loss 
than the ones presented in this study. 

\subsection{Equation of state}
\label{sect:eqstate}

In this study, we used very simple equations of state: isothermal 
(\eq{eq:state}) or adiabatic (\eq{eq:pressure}). The effect of the equation 
of state can be appreciated by seeing the evolution of the Solberg-H{\o}iland criterion 
in isothermal and adiabatic simulations. In the former, it is the epicyclic 
frequency that brings $\kappa^2+N^2$ to negative values, while in the latter the 
criterion is broken mainly by the Brunt-V\"ais\"al\"a frequency. 
The excitation of the Rossby wave instability is greatly favored in the presence 
of a strong entropy gradient, and made more difficult (yet not impossible) as the 
disk approaches isothermality. Therefore, an accurate modeling of the energy 
budget - solving for radiative cooling and turbulent heating -, is something to 
pursue in order to more realistically address the evolution of the RWI and the issue 
of planet formation inside the vortices that constitute its saturated state. 

\subsection{Aerodynamics of the embryo}
\label{sect:aerodynamics}

The aerodynamics of a super-particle is controlled by the radius $a_\bullet$ of 
the individual rocks. This means that even though the ensemble of 
rocks has the same position and velocity, there is still 
space between them and therefore they have contact with the gaseous nebula 
through all their surface area. 

However, after gravitational collapse occurs, the solids are not any longer 
an ensemble of pebbles and boulders with free space between them, but a single massive 
object of planetary dimensions. This leads to a radical change in the 
aerodynamical properties. Yet, in our simulations, we still consider the collapsed body as an 
ensemble of super-particles, with the aerodynamical properties of individual 
pebbles and boulders. This is certainly a limitation of the model. 

For a large planet, the correct treatment would be to consider that, after collapse,
we leave the regime of particle-gas Stokes drag and enter 
planet-disk interaction by gravitational friction (type I migration). In the 
solar nebula the two drags have similar strength for bodies of 100\,km.  
As we solve for the self-gravity of the gas, the latter is included in our model, 
albeit limited by the resolution of the grid. The fact that we keep using 
Epstein-Stokes drag on the super-particles after collapse might make an 
embryo more stable, especially in view of the very effective 
dynamical cooling provided by the drag force (\fig{fig:collapse}). 
Substituting collapsed clusters by a sink particle that feels the gas 
gravity but not the gas drag is a possible solution, but also has 
caveats on its own. The evolution of sink particles depends too much on 
artificial numerical parameters such as the accretion radius. Furthermore, 
a sink particle does not suffer tidal effects, which we showed to be 
non-negligible.

\subsection{Coagulation and Fragmentation}
\label{sect:coagulation-fragmentation}

As dust particles are drawn together,  
electromagnetic interactions occur at their surface, 
causing sticking under favorable conditions. Brauer et al. (2008b) 
show that density enhancements like the ones we see -
where matter accumulates due to a discontinuity in viscosity -, dramatically 
favor coagulation. As particles are drawn together and the relative velocities are 
reduced, growth by coagulation occurs for a range of mass accretion rate $\mdot$ and 
threshold fragmentation velocity $v_{\rm ft}$. They find that the meter size barrier
can be breached for mass accretion rates up to $\mdot$=$\ttimes{-8}$$M_\Sun/{\rm yr}$ 
(for $v_{\rm ft}$=10\,\mps) and threshold fragmentation velocities no 
less than $v_{\rm ft}$=5\,\mps (for $\mdot$=$\xtimes{8}{-9}$\,$M_\Sun/{\rm yr}$). 

This raises the possibility that even before the RWI excites the vortices, 
coagulation will have depleted the population of centimeter and meter sized 
objects onto bodies that are too large for our proposed mechanism to occur efficiently. 
As we see, it is preferentially the 10 and 30\,cm sized particles that 
concentrate into planetary embryos.

The timescale for coagulation, however, is much longer than the time-scale 
for gravitational collapse. We see growth to Mars size taking place 
in only five orbits ($\approx$60 yr), while growth by coagulation 
from meter to kilometer size occurs on timescales of a few thousand years 
according to Brauer et al. (2008b). On the other hand, it could as 
well be that the favorable environment provided by the vortices act as to 
speed up coagulation even further. This, of course, is not bad. Growth beyond 
the preferred size will lead to decoupling from the gas and ejection from the 
vortex that, in the end, behaves as a planetesimal factory.

Fragmentation is an important piece of reality that we did not include in our 
 model. Nevertheless, we showed in Sect~4.5 that the majority of the particles 
were never involved in collisions with speeds greater than 1\mps. These are 
of course very good news for planet formation. However, we feel the need to 
stress that the time interval between snapshots in 
\fig{fig:collision}a is of whole orbits (totaling 200 snapshots). 
The number of high-speed 
impacts could be greater had we checked the collision speeds at every 
time-step instead. Although desirable, this would have made the simulations 
computationally very expensive since it must be done in runtime. The result 
of \fig{fig:collision}b should therefore be taken only as further evidence 
for low collisional speeds inside vortices, not as conclusive proof of it. 
Carballido et al. (2008) further point that the low collisional 
speeds at low separation may be unrealistic. This is because the particles 
couple to the smallest eddies, whose size is a function of the mesh 
Reynolds number. These authors find average collisional speeds of 0.05$c_s$ 
for particles of stopping time $\Omega_K\tau_f$=0.2 ($a_\bullet$$\approx$15\,cm 
in our models), but notice a sharp decrease of the collisional 
speeds towards smaller separations. In our simulations, we are considering 
encounters that happen inside a grid cell, where we do not resolve the 
velocity field, so this may indeed be the reason behind the low collisional velocities 
we find. However, we point that there is a difference between the simulations of 
Carballido et al. (2008) and those presented in this paper. They considered 
particle concentrations in the transient pressure maxima of the 
turbulence generated by the MRI, whereas we consider particle concentrations 
within long lived anticyclonic vortices. As vortex structures tend to 
merge and grow, there is less power available at the smallest scales when 
compared to MRI turbulence.

\subsection{Disk mass}
\label{sect:diskmass}

We stated in Sect~2.2 that the constrain of having 30 Earth masses 
of solid in the inner disk, together with the interstellar dust-to-gas 
ratio of $\epsilon$=$\ttimes{-2}$, and a surface density slope of -0.5 implies 
0.01 solar mass within the modeled range. This disk can be considered 
massive since the shallow slope then implies a massive outer disk: 
0.3 solar mass within 100\,AU. The actual mass is somewhat lower 
because the disk goes isothermal after a given radius. If the temperature 
at 5.2\,AU is 100K, it will have fallen to 10\,K at 50\,AU and below the 
cosmic microwave background temperature at 200\,AU. As the surrounding 
temperatures of the molecular cloud are of $\sim$10-20\,K, the $1/r$ fall of the 
temperature has to stop at some point. Boss (2002) and Pinotti et 
al. (2005) account for this by using a temperature profile of 
$T = T_0 r^{-q_T} + T_B$, where the parameter 
$T_B$ works as a background temperature (see also 
Papaloizou \& Terquem 1999), leading to an essentially 
isothermal profile after a certain distance. The 
constancy of the mass accretion rate then dictates that the slope of 
the surface density has to change accordingly, to -1.5. This, in turn, leads 
to a lower disk mass. Taking into account this isothermal profile
in the very outer disk, using $T_B$=10\,K, and
the corresponding change of slope in $\varSigma$, the disk sets at a
constant value of $Q$=1 after 50\,AU, with a mass of 0.25 solar mass at 100\,
AU. If $T_B$=20\,K, the mass decreases to 0.16 $M_\odot$ at 100\,AU, with 
$Q$=3 after 25\,AU. 

Observational studies in contrast obtain a distribution of 
disk masses with median of $\xtimes{5}{-3}\,M_\odot$ (Andrews \& 
Williams 2005). However, considering 
an interstellar dust-to-gas ratio of $\ttimes{-2}$, such disks 
contain only 15 \mearthp of solid material, and cannot form the 
solar system. We take it as an indication that or the masses are 
underestimated (Hartmann 2008) or that, as these 
observations correspond to relatively older disks (1-4\,Myr), the 
gas has already dissipated significantly (Hueso \& Guillot 2005). 
Our model then should be representative of young disks, therefore 
corresponding to an early formation of relatively massive 
planetary embryos.

\section{Summary and conclusions}
\label{sect:conclusions}

We have undertaken simulations of disks 
of gas and solids, where the solids are represented by Lagrangian particles 
of radii 1, 10, 30, and 100\,cm. We show that in the borders 
of the dead zone, where the accretion flow stalls due a difference in 
turbulent viscosity, rapid gravitational collapse of the particles into 
protoplanets occurs within the vortices that form due to the excitation 
of the Rossby wave instability (Lovelace et al. 1999). As shown in LJKP08, 
over 300 gravitationally bound planetary embryos were formed, 20 of them being 
more massive than Mars. The mass spectrum follows a power law of index 
$-$2.3$\pm$0.1 in the interval $-$2.0$<$$\log$($M$/\mearth)$<$ $-$1.2.

Although for the main results of this study we have used sharp viscosity jumps
to model the transition between the active and dead zones, we show that the RWI
is excited up to viscosity jumps as smooth as $\Delta{r}$=$2H$ where $H$ is the
pressure scale height. For this conclusion, we used the Solberg-H{\o}iland
criterion as providing a conservative estimate of whether the RWI is excited.
The consequence of increasing the width of the viscosity jumps seems to be that
the threshold of instability is reached at increasingly longer
times. It only takes five orbits with $\Delta{r}$=$0.2H$ (the usual width used
in the models presented in this paper), but takes 350 orbits for
$\Delta{r}$=$2H$. We also assessed if the vortices would survive in the more
realistic environment of a turbulent disk by making the location of the
viscosity shift oscillate with an amplitude typical of the scale length of MRI
turbulence over the period of a Keplerian revolution. We find that this has
little effect on the excitation of the RWI and saturation into vortices. 

We model the solid phase with Lagrangian superparticles representing
physical pebbles and rocks of different size (1, 10, 30, and 100\,cm). As these
particles are subject to different gas drag, an aerodynamical sorting by size
takes place within the vortices. The consequence of this is that the first
bound structures are formed of single particles species. This is a very
interesting result, since it is an observed fact that planetesimals are formed
of similar-sized building blocks (Scott \& Krot 2005). These building blocks
seem to be sub-mm sized grains, but different nebula parameters could work as
to trap smaller particles. Youdin (2008) also points that the stationary point
of a particle trapped in vortical motion is shifted azimuthally with respect to
the eye, according to its radius $a_\bullet$. We indeed see that clumps of
particles of different size, which collapse into different embryos inside the
same vortex, usually retain significant azimuthal displacements between each
other for long times instead of forming a single, more massive, embryo at the
vortex eye. This may or may not be a result of the size-dependent azimuthally
shifted stationary points of  Youdin (2008).

A collapsed embryo is observed to be very compact. The compactness is mainly
provided by the drag force, which makes the system very dissipative (dynamical
cooling by gravity alone works on much longer timescales). Collapse towards
``infinite'' density is seen to occur in some cases, with most of the particles
occupying the same position in space (limited by numerical precision). In the
specific case when the particles dominate the time-step, the Courant condition
leads a particle to overshoot the center of the mass, so that it executes
oscillations about it, which in turn leads to a finite rms radius. We also
observe that a clump of particles is susceptible to the 
disrupting effects of ram pressure erosion and gravitational tides
from the gas. Both effects are proportional to the local gas-to-solids density 
ratio. When the vortices in the inner border of the dead zone undergo the transition
from the $m$=3 to the $m$=2 mode, their surface density increases, with drastic
consequences for the embryos within them. The most massive embryo by that time,
a protoplanet 6.7 times the mass of Mars, mostly formed of $a_\bullet$=30\,cm
particles, was split into two smaller objects, of 5.9 and 0.8 $M_{\rm Mars}$,
due to the action of the gas tides. The fate of the embryos formed mostly of
10\,cm was more dramatic. As the 10\,cm particles experience stronger drag
forces, the ram pressure is also stronger. During the mode transition, the
combined effects of tides and erosion completely obliterated these embryos,
leaving extended arcs of particles that did not reaccumulate until the end of
the simulation. We anticipate that this effect will be very reduced in 3D
simulations. In 2D simulations, the ratio of the vertically integrated solids
density to the gas column density $\varSigma_p/\varSigma_g$ never gets much
above unity even for the most massive embryo. In contrast, the ratio of the
bulk density of solids to the volume gas density $\rho_p$/$\rho_g$ is greatly
increased in the midplane of 3D disks due to sedimentation.

We also observe that the solids modify the evolution of the RWI. We are drawn
to this conclusion because a simulation without the backreaction of the drag
force from the particles onto the gas developed vortices at later times when
compared to the ones that include particle feedback. We stress that this is not
due to the streaming instability, since the solids-to-gas ratio was much lower
than the value needed to excite it ($\rho_p/\rho_g$ $\apprge$ 1). Instead, it
is more likely that the backreaction of the drag force contributes
non-negligibly to the dispersion relation of the RWI. Self-gravity is also
seen to play a role on modifying the evolution of the turbulence. We observe
that in simulations without self-gravity, the disk tends to show less vortices
at later times. In a simulation where we switched off the self-gravity after
the vortices had developed, the dominant $m$=$4$ mode in the outer edge of the
dead zone rapidly turned into a $m$=3 mode. The vortices in the inner edge also
quickly turned from a dominant $m$=3 mode to displaying a $m$=2 mode instead.
We measured the Toomre $Q$ parameter and found that the vortices have 
$Q\approx 1$. This constitutes further evidence that in the presence of 
self-gravity, vortex growth is not only limited by the pressure scale height 
but also by the Jeans length (Mamatsashvili \& Rice 2009).

An important finding in this paper is that under vortex trapping, the
collisional
speeds between particles are greatly reduced. We measured the collisional
velocity history of every particle that is bound to the most massive embryo at
the end of the simulation, and found that the vast majority of them never
experienced close encounters at speeds greater than 1 \mps. This is well below
the fragmentation threshold and lends support to the
long-held idea that vortices provide a superbly favorable environment for
planetary growth. Growth by coagulation beyond the optimal size for planet
formation is also avoided because the timescales for coagulation are much
longer than the rapid timescale for gravitational collapse witnessed in our
models. This does not exclude the possibility that
coagulation itself is sped up within a vortex. In this case, the vortex will
behave as a planetesimal factory, quickly producing kilometer sized bodies that
leave the vortex due to their decoupling from the gas.  This, as noted by Klahr
\& Bodenheimer (2006) is very positive for planet formation, since the formed
planetesimals are then scattered through the disk, where they can be used to
form planets independently of a vortex. Even though it implies that
we are facing the comfortable position of a win-win situation for planet
formation, one has to decide which process (planet formation by direct
gravitational collapse or planetesimal formation by fast coagulation) is
getting the upper hand inside the Rossby vortices. A definite answer to this
question can only be drawn from a simulation that includes the processes of
coagulation/fragmentation. Unfortunately, inclusion of sophisticated
coagulation/fragmentation models such as that of Brauer et al.  (2007) would
render a hydrodynamical simulation overly expensive. A possible alternative
would be a Monte Carlo model of dust coagulation, such as the one recently 
developed by Ormel et al. (2007). A further development of the Monte 
Carlo technique is described in Zsom \& Dullemond (2008). The main difference 
between the two models is that while Ormel et al. (2007) simulate coagulation 
between real dust particles, Zsom \& Dullemond (2008) use superparticles to 
model coagulation and fragmentation. Therefore the latter one is more suitable 
for hydrodynamical simulations such as ours and simple estimations show that 
this model could be adapted to a hydro model with no prohibitive overhead.

We reiterate that the models presented suffer from a number of limitations,
detailed in Sect~\ref{sect:limitations}. Some of them, like refining the
particle mass resolution to the individual pebbles and rocks, are beyond the
capabilities of the current generation of computer models. Others, however,
such as inclusion of detailed thermal physics, could be tackled with relatively
little effort. We urge researchers active on the field to consider these
problems. It is our hope that a coherent picture of planet formation in
the magnetically dead zones of accretion disks shall emerge as a result
of it.

\begin{acknowledgements}
Simulations were performed at the PIA cluster of the Max-Planck-Institut
f{\"u}r Astronomie and on the Uppsala Multidisciplinary Center for Advanced
Computational Science (UPPMAX). This research has been supported in 
part by the Deutsche Forschungsgemeinschaft DFG through grant DFG 
Forschergruppe 759 ``The Formation of Planets: The Critical First 
Growth Phase''. A. Zsom acknowledges support by the IMPRS for 
Astronomy \& Cosmic Physics at the University of Heidelberg.
\end{acknowledgements}

\Online

\begin{appendix}

\section{Resolution study}
\label{app:resolution}

\begin{figure}
  \begin{center}
    \includegraphics[width=\hfwidthsingle]{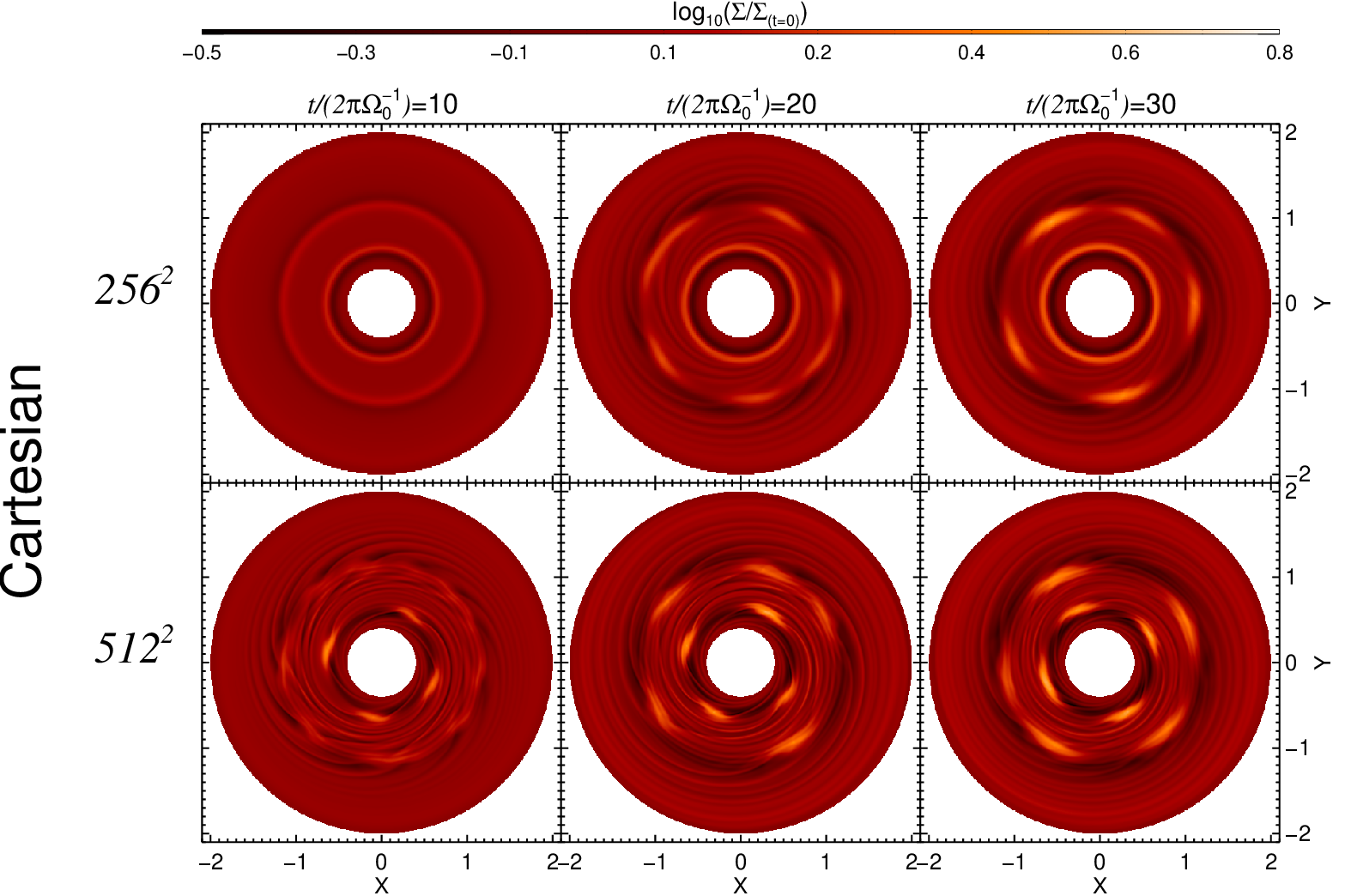}
    \includegraphics[width=\hfwidthsingle]{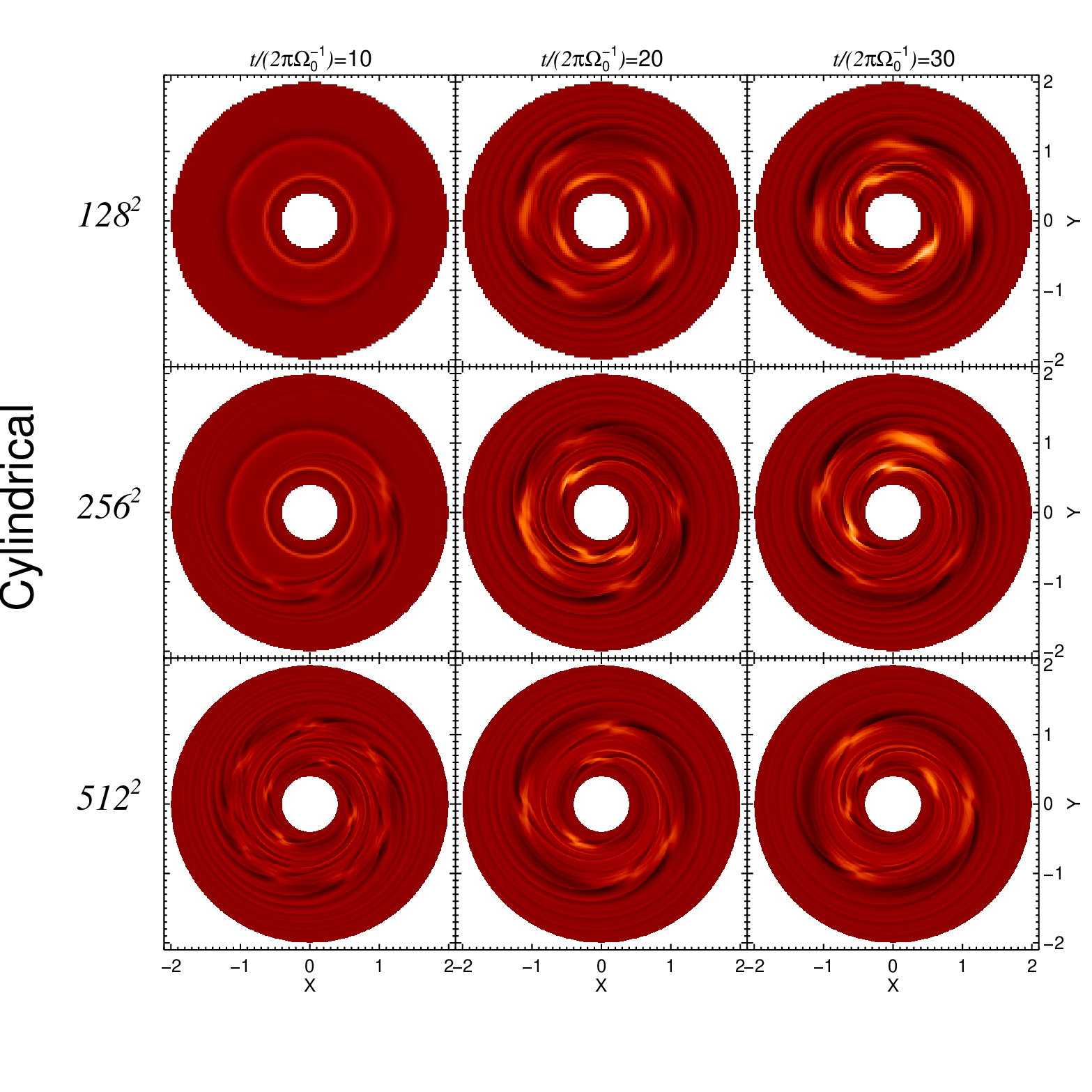}
  \end{center}
  \caption[]{The development of the dead zone vortices in different resolution 
and grid geometry. With increased resolution the RWI is excited increasingly earlier. 
In general the vortices in cylindrical runs look sharper than in the 
Cartesian ones, due to the better azimuthal resolution. The Cartesian 
run of middle resolution (256$^{\rm 2}$) has too coarse azimuthal 
resolution in the inner disk, and only developed vortices in the inner 
edge of the dead zone at later times ($\approx$ 40 orbits). Apart from 
these differences, the vortices look remarkably similar, having nearly 
the same peak density and same vorticity.}
  \label{fig:resolution}
\end{figure}

In \fig{fig:resolution} we present a resolution study of our models. The upper 
panels show Cartesian models, including self-gravity and solids, while the 
lower panels show cylindrical models with dustless non-selfgravitating gas. 
Both are shown in selected snapshots at 10, 20, and 30 orbits.

In the Cartesian runs, the vortices in the high-resolution run (512$^{\rm 2}$) 
are excited earlier than in the middle resolution run (256$^{\rm 2}$).
At later times, the vortices in the high-resolution run also 
appear sharper. While in the two runs the vortices in the outer edge 
of the dead zone look remarkably similar, the inner edge displays 
very different behavior. Up to 30 orbits the middle resolution run has not 
shown signs of prominent vortices. In contrast, the high resolution run 
had the 
inner edge developing vortices as early as 10 orbits. This is not only 
due to the high resolution run having a wider inertial range, but also 
because a flow with cylindrical symmetry is more coarsely resolved 
near the origin in a Cartesian grid. Because of this, the most unstable 
modes are under-resolved in the inner disk of the middle resolution run. 
A Cartesian run with low resolution (128$^{\rm 2}$, not shown) did not 
develop vortices even in the outer disk by the same time. At 30 orbits, 
the density inside the vortices peak at similar values, $\varSigma/\varSigma_0$=3.3 
and 3.7 for the middle and high resolution runs, respectively. 

The cylindrical runs also illustrate the small amount of differences between 
the vortices in different runs. With better azimuthal resolution, the 
vortices are excited even in the 128$^{\rm 2}$ run, and in both runs they 
peak with surface density $\varSigma/\varSigma_0$=4.6. In fact, the main effect 
of resolution appears to be that, as it increases, the RWI is excited 
increasingly earlier. The high-resolution run displays weaker vortices 
than the others because in this case we were forced to use shock 
viscosity. 

\section{Emulating turbulent motions} 
\label{app:vary-nu-time}

\begin{figure*}
  \begin{center}
    \includegraphics[width=.85\textwidth]{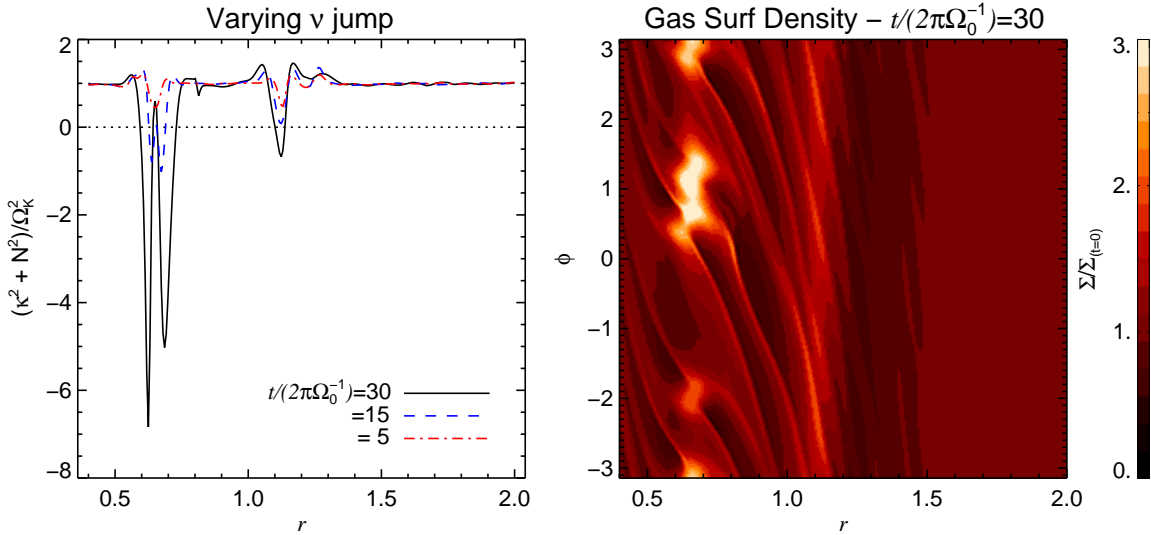}
  \end{center}
  \caption[]{Evolution of the RWI with a time-varying viscosity jump 
as specified by \Eq{eq:viscosity} and \Eqss{eq:r1vary}{eq:r2vary}.
The panel on the left-hand side measures the violation of the 
Solberg-H{\o}iland criterion. The right-hand side panel shows the 
appearance of the disk at 30 orbits. Both the inner and outer edge quickly 
reach the threshold of instability (left panel). At 30 orbits, the inner 
edge already reached a saturated state and launched vortices (right panel).}
  \label{fig:varynu}
\end{figure*}

In this study, we considered the dead zone to be represented by a static 
viscosity profile. In a more realistic scenario, turbulent motions caused 
by the MRI and 
variations in the coupling between the magnetic field and the plasma will 
give rise to a turbulent resistivity. This is expected to cause the border 
of the dead zone to vary in space and time, with implications for the 
evolution of the RWI.

To assess the impact of space and time variability of the edges  
of the dead zone, we model the viscous jumps using $\Delta{r}$=$0.01$ but make 
the jumps shift radially in time by substituting $r_1$ and 
$r_2$ in \Eq{eq:viscosity} by 

\begin{eqnarray}
  r_1(t) &\rightarrow& r_1\left[1 + h\sin\left(\varOmega_K(r_1)t\right)\right]
  \label{eq:r1vary}\\ 
  r_2(t) &\rightarrow& r_2\left[1 + h\sin\left(\varOmega_K(r_2)t\right)\right]
  \label{eq:r2vary}
\end{eqnarray}where $h$=$H/r$ is the aspect ratio. So, the location shifts by 
two scale heights over a Keplerian revolution. The results are shown 
in \fig{fig:varynu}, where we show the appearance of the disk 
at 30 orbits and the azimuthal average of $(\kappa^2+N^2)/\varOmega_K^2$ 
(in the same 2D model, as opposed to 1D as in Sect.~\ref{sect:sharpness}).

In this simulation, the RWI 
is still excited and vortices are launched. The main difference when compared 
to simulations with static profiles is that the instability takes more time 
to violate the Solberg-H{\o}iland criterion, $\approx$ 10 orbits, compared 
to 5 in the static case. This is due to the fact that the shifting viscosity 
jump smears the pressure maximum, so the amplitude of the pressure jump 
is shorter and the width is larger than in the static case. 

\section{Onset of erosion}
\label{app:erosion}

\begin{figure*}
  \begin{center}
    \includegraphics[width=.8\textwidth]{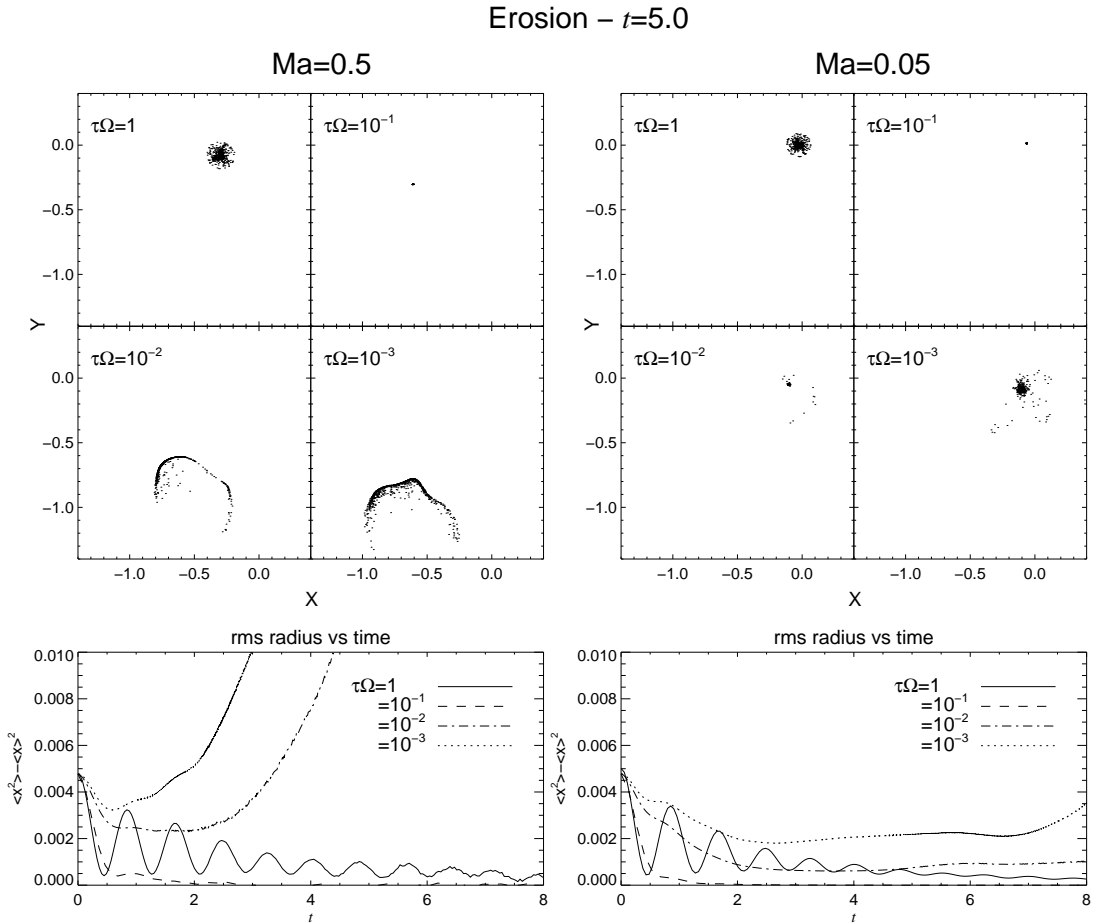}
  \end{center}
  \caption[]{A clump of 1000 massive particles moving against headwinds 
of $\Ma$=0.5 and $\Ma$=0.05, for different values of the friction time $\tau\varOmega$. 
In the case of $\Ma$=0.5, self-gravity cannot hold the clump together for 
$\tau\varOmega \apprle 10^{-2}$. In our simulations it corresponds to 1\,cm sized particles, 
approximately. Clumps formed of larger particles to not experience erosion. For 
the more subsonic motion, the effect of ram pressure is lower so the clump of 
$\tau\varOmega$=$10^{-2}$ is more stable against erosion. The case of 
$\tau\varOmega$=1.0 takes longer to contract because of the weaker drag force, which 
provides less dynamical cooling than in the case with $\tau\varOmega$=0.1. It eventually 
shrinks, as seen in the time series (bottom plot).}
  \label{fig:erosion}
\end{figure*}

According to \eq{eq:weber}, a tight distribution of particles under Epstein drag 
should suffer erosion if

\begin{equation}
  {\rm We}_G = \frac{\Ma c_s\varOmega}{(\tau\varOmega)\pi G\varSigma_p} \geq 1
\end{equation} 

In this appendix, we perform numerical simulations to test the validity of this 
condition. We model a clump of $\ttimes{3}$ particles suffering a strong headwind from 
the gas ($\Ma$=1/2). The blob of particles is initially set as a tight Gaussian distribution 
about the center of the grid, with surface density peak of $\varSigma_p/\varSigma_0=2.31$. In units where 
$G$=$c_s$=$\varSigma_0$=$\varOmega_0$=1, the initial gravitational Weber number at the 
surface of the blob is therefore ${\rm We}_G \simeq \xtimes{7}{-2}/(\tau\varOmega)$. 

We plot in \fig{fig:erosion} the evolution of clumps for four different values of 
$\tau\varOmega$. For $\tau\varOmega$=1.0 and $\tau\varOmega$=0.1, ${\rm We}_G$ is less than 1 so the 
clump is stable against ram pressure and contracts. The other clumps ($\tau\varOmega$=0.01 
and $\tau\varOmega$=0.001) have ${\rm We}_G$ above unit, and experience intense erosion. 
We also considered a flow of Mach number $\Ma$=0.05. In this case the initial 
gravitational Weber number is ${\rm We}_G \simeq \xtimes{7}{-3}/(\tau\varOmega)$, and the 
clumps of smaller particles are supposed to be more stable. Indeed, this is what we 
see in the figure. The clump of particles of $\tau\varOmega$=0.01 is now marginally 
stable and contracts, experiencing much less erosion than in the $\Ma$=0.5 case.

We would like to draw attention to an interesting feature of \fig{fig:erosion}. 
The cases of $\tau\varOmega$=1 and $\tau\varOmega$=0.1 (upper panels) 
provide yet another perspective for the action 
of drag force cooling. Both clumps are 
stable against erosion but the clump of $\tau\varOmega$=0.1 has shrunk
considerably more than the clump of $\tau\varOmega$=1.0, which looks very 
extended. What is happening is that the clump with $\tau\varOmega$=1 is 
too weakly coupled to the gas and therefore takes longer to collapse, as 
seen in the time series of the rms position (\fig{fig:erosion}, lower panels).

\end{appendix}

\end{document}